\documentclass[11pt,journal,final,letterpaper,onecolumn]{article}
\usepackage[top=1in, bottom=1in, left=1in, right=1in]{geometry}

\usepackage{setspace}
\usepackage[utf8]{inputenc} %
\usepackage[T1]{fontenc} %
\usepackage{hyperref} %
\usepackage{url} %
\usepackage{booktabs} %
\usepackage{amsfonts} 
\usepackage{amsmath}%
\usepackage{nicefrac} %
\usepackage{microtype} %
\usepackage{bm} %

 \usepackage{graphicx}

 \usepackage{algorithm}

\title{\Large{Graph Gamma Process Generalized Linear Dynamical Systems}}

 \usepackage{hyperref}
 \usepackage{nccmath}
 \usepackage{xcolor} 
 \usepackage{booktabs} 
\usepackage{hyperref}
 \usepackage{amssymb,amsmath,amsthm,enumitem}
\usepackage{algorithm}
\usepackage[noend]{algpseudocode}

\usepackage[round]{natbib}
\usepackage{cite}

\usepackage[labelformat=simple]{subcaption}

\newcommand*{\rom}[1]{\expandafter\@slowromancap\romannumeral #1@}
\newcommand{\beq}{\vspace{0mm}\begin{equation}}
\newcommand{\eeq}{\vspace{0mm}\end{equation}}
\newcommand{\beqs}{\vspace{0mm}\begin{eqnarray}}
\newcommand{\eeqs}{\vspace{0mm}\end{eqnarray}}
\newcommand{\barr}{\begin{array}}
\newcommand{\earr}{\end{array}}
\newcommand{\Amat}[0]{{{\bf A}}}

\newcommand{\Dmat}{{\bf D}}
\newcommand{\Emat}[0]{{{\bf E}}}

\newcommand{\Hmat}[0]{{{\bf H}} }
\newcommand{\Imat}{{\bf I}}

\newcommand{\Mmat}{{\bf M}}

\newcommand{\Rmat}{{\bf R}}

\newcommand{\Wmat}[0]{{{\bf W}}}

\newcommand{\Zmat}{{\bf Z}}

\newcommand{\dv}{\boldsymbol{d}}

\newcommand{\mv}[0]{{\boldsymbol{m}}}

\newcommand{\xv}{\boldsymbol{x}}
\newcommand{\yv}{\boldsymbol{y}}

\newcommand{\cdotv}{\boldsymbol{\cdot}}

\newcommand{\Thetamat}{\boldsymbol{\Theta}}

\newcommand{\Lambdamat}{\boldsymbol{\Lambda}}

\newcommand{\bas}[1]{\begin{align*}#1\end{align*}}
\newcommand{\ba}[1]{\begin{align}#1\end{align}}

\newcommand{\Sigmamat}[0]{{\boldsymbol{\Sigma}}}

\newcommand{\Phimat}{\boldsymbol{\Phi}}
\newcommand{\Psimat}{\boldsymbol{\Psi}}
\newcommand{\Omegamat}[0]{{\boldsymbol{\Omega}}}

\newcommand{\thetav}{\boldsymbol{\theta}}

\newcommand{\lambdav}[0]{{\boldsymbol{\lambda}}}
\newcommand{\muv}[0]{{\boldsymbol{\mu}}}

\newcommand{\xiv}[0]{{\boldsymbol{\xi}} }

\newcommand{\psiv}{\boldsymbol{\psi}}

\newcommand{\E}{\mathbb{E}}

\newcommand{\given}{\,|\,}

\newtheorem{thm}{Theorem} %

\newtheorem{lem}[thm]{Lemma}

\author{%
  Rahi Kalantari,\quad Mingyuan Zhou\\
  The University of Texas at Austin\\
  Austin, TX 78712}

\begin{document}

\maketitle 

\begin{abstract}
We introduce graph gamma process (GGP) linear dynamical systems to model  real-valued multivariate time series. For temporal pattern discovery, the latent representation under the model is used to decompose the time series into a parsimonious set of multivariate sub-sequences. In each sub-sequence, different data dimensions often share similar temporal patterns but may exhibit distinct magnitudes, and hence allowing the superposition of all sub-sequences to exhibit diverse behaviors at different data dimensions. We further generalize the proposed model by replacing the Gaussian observation layer with the negative binomial distribution to model multivariate count time series. Generated from the proposed GGP is an infinite dimensional directed sparse random graph, which is constructed by taking the logical OR operation of countably infinite binary adjacency matrices that share the same set of countably infinite nodes. Each of these adjacency matrices is associated with a weight to indicate its activation strength, and places a finite number of edges between a finite subset of nodes belonging to the same node community.  We use the generated random graph, whose number of nonzero-degree nodes is finite, to define both the sparsity pattern and dimension of the latent state transition matrix of a (generalized) linear dynamical system. The activation strength of each node community relative to the overall activation strength is used to extract a multivariate sub-sequence, revealing the data pattern captured by the corresponding community. On both synthetic and real-world time series, the proposed nonparametric Bayesian dynamic models, which are initialized at random,  consistently exhibit  good  predictive performance in comparison to a variety of baseline models, revealing interpretable latent state transition patterns and decomposing the time series into distinctly behaved sub-sequences.

\end{abstract}

\section{Introduction} \label{intro}

Linear dynamical systems (LDSs) have been widely used to model real-valued time series
\citep{Kalman,West:1997:BFD:261170,ghahramani1999learning,Ljung}, with diverse applications such as financial time series analysis \citep{carvalho2007simulation} and movement trajectory modeling
\citep{gao2016linear,zhang2017sensor}. 
They have become standard tools in
optimal filtering, smoothing, and control \citep{imani2018particle, hardt2018gradient, koyama2018projection}.   An LDS consists of two main blocks, including an observation model, which 
assumes that the observations are translated from their latent states via 
a linear system with added Gaussian noise, and a transition block, which is represented by a Markov chain that linearly transforms a latent state from time $t-1$ to time $t$  with added Gaussian noise. The transition block plays an important role in capturing the underlying dynamics of the data. An LDS, which has limited representation power due to its linear assumption, allows one to examine the temporal trajectory of each latent dimension to understand the role played by the corresponding latent factor. While it is often considered to be simple to interpret, its interpretability often quickly deteriorates  as its latent state dimension increases. 

To enhance the representation power of LDSs, in particular, to model non-linear behaviors of the time series  and improve their interpretability, 
 one may consider
 switching LDSs 
 \citep{FoxSLDS09,pmlr-v54-linderman17a,nassar2018tree}, which learn how to divide the time series into separate temporal segments and fit them by switching between different LDSs. 
Important parameters include the number of different LDSs, their latent state dimensions, and the transition mechanism from one LDS to another. 
 While nonparametric Bayesian techniques have been applied to switching LDSs to learn the number of LDSs that is needed, the latent state dimensions often stay as tuning parameters to be set  \citep{FoxSLDS09,nassar2018tree}. Moreover, switching LDSs do not allow different LDSs to share latent states, making it difficult to capture smooth transitions between different temporal patterns, and false positives/negatives and delays in detecting the switching points will also compromise their performance. In addition, %
 existing optimal smoothing and filtering  techniques  developed for LDSs, such as Kalman filtering \citep{Kalman}, require non-trivial modifications before being able to be applied to switching LDSs \citep{murphy1998switching}.

 Moving beyond switching LDSs where different LDSs neither share their latent states nor overlap in time, we propose the graph gamma process (GGP) LDS that encourages forming multiple LDSs that can share their latent states and co-occur at the same time.  GGP-LDS uses a flexible combination of multiple LDSs to fit the observation at any given  time point, allowing smooth transitions between different dynamical patterns across time. A notable %
 feature of GGP-LDS is that existing optimal filtering and smoothing techniques developed for a canonical LDS can be readily applied to GGP-LDS without any modification. Therefore, GGP-LDS can serve as a plug-in replacement of the LDS in an existing system.
 The introduced nonparametric Bayesian construction in GGP-LDS will support $S$ latent states that are shared by $K$ different types of LDSs, each of which is characterized by its own pattern of activation probabilities imposed on a $S\times S$ sparse state-transition matrix, and allow both $S$ and $K$ to grow without bound. 
 This unique  construction is realized by modeling
 the sparsity structure of the $S\times S$ state-transition matrix as the adjacency matrix of a directed random graph, which is resulted from the logical OR operation over %
 $K$ latent binary adjacency matrices, each of which is drawn according to the interaction strengths between the states (nodes) of a type of LDS (node community).  Each latent binary adjacency matrix can be considered as a specific type  of state community, which describes a particular type of relationship between latent states, $i.e.$, for a given state of the current time, which states of the previous states will directly influence it, and which states of the future time will be directly influenced by it. While a latent state is associated with all communities, the association strengths can clearly differ. Note that the sparsity pattern of the state-transition matrix  is determined by the logical OR of these community-specific binary adjacency matrices. Therefore, to facilitate interpretation and visualization,  one can hard assign a state to a community whose binary adjacency matrix  best explains how this state is being influenced by the states of the previous time, or to a community that best explains how this state is influencing the states of the next time. 

GGP-LDS 
allows approximating complex nonlinear dynamics by 
activating a certain combination of communities to model a particular type of linear dynamics at any given time, and using smooth transitions between overlapping communities %
to model smooth transitions between distinct linear dynamics. 
The characteristics of each community can be visualized by reconstructing the observations using the inferred latent representation and a community-specific reweighted latent state-transition matrix,  where the weights are determined by the  activation strength of that community relative to the combined activation strength of all communities.
In addition to modeling real-valued time series, we show how to generalize %
the observation layer of  GGP-LDS %
to model overdispersed count time series.
It is noteworthy to mention that while the LDS \citep{kalman1963mathematical} has been chosen as the transition and observation model of GGP-LDS, the proposed  GGP can potentially be applied to many other nonlinear systems that have a latent state transition module  \citep{johnson2016composing}.

\section{Nonparametric Bayesian Modeling}
\label{gen_inst}
For LDSs, 
let us denote $\yv_t\in\mathbb{R}^V$ and $\xv_t\in\mathbb{R}^S$ as the observed data and latent state vectors, respectively, at time $t\in\{1,\ldots,T\}$, $\Dmat=(\dv_1,\ldots,\dv_S)\in\mathbb{R}^{V\times S}$ as the observation factor loading matrix, and both $\Phimat\in\mathbb{R}^{V\times V}$ and $\Lambdamat =\mbox{diag}(\lambda_1,\ldots,\lambda_S)$ %
as precision (inverse covariance) matrices. Inspired by \citet{Kalantari2018nonparametric}, we first modify the usual LDS hierarchical model by utilizing a
spike-and-slab construction \citep{mitchell1988bayesian,ishwaran2005spike,Mingyuan09}, %
which imposes binary  mask $\Zmat \in \{0,1\}^{S \times S}$ element-wise on the real-valued latent state transition matrix $\Wmat\in \mathbb{R}^{S \times S}$~as 
$$ \yv_t\sim\mathcal{N} (\Dmat \xv_{t},\Phimat^{-1}), 
~\xv_t\sim\mathcal{N} \left(\left( \Wmat \odot \Zmat \right) \xv_{t-1},\Lambdamat^{-1}\right).$$

For count observations, we modify the Gaussian distribution based observation layer with the negative binomial (NB) distribution %
by letting
$
\small \yv_t\sim\mbox{NB} \left(\eta,%
{\sigma(\Dmat\xv_t)
}
\right), ~\eta\sim\mbox{Gamma}(\alpha_{\eta},1/\beta_{\eta}),
~\xv_t\sim\mathcal{N} \left(\left( \Wmat \odot \Zmat \right) \xv_{t-1},\Lambdamat^{-1}\right),$ where $\sigma(x):={1}/({1-e^{-x}})$ is the sigmoid function.  %

The $K\times K$ latent state-transition matrix $\Wmat\odot \Zmat$, in particular, the sparsity structure of~$\Zmat$, plays an important role in determining the dynamical behaviors %
of the model. First, the nonzero locations in $\Zmat$ determine the temporal dependencies between the latent states \citep{Kalantari2018nonparametric}. For example, if $z_{ij}$, the $(i,j)$th element of $\Zmat$, is zero,  then at time $t$, $x_{ti}$ will be independent of $x_{(t-1)j}$, and $x_{tj}$ will not influence $x_{(t+1)i}$. Thus in what follows, we consider that there is a directed link (edge) from states (nodes) $j$ to $i$ if $z_{ij}=1$.

Second, viewing $\Zmat$ as the adjacency matrix of a directed random graph and the states of the LDS as the graph nodes, we may introduce inductive bias to encourage its nodes to be formed into overlapping communities, reflected by overlapping dense blocks along the diagonal of the adjacency matrix after appropriately rearranging the orders of the nodes. We may then view each community as an LDS,
which forms its own state-transition matrix, using a submatrix of $\Wmat\odot \Zmat$, to model the transitions between the corresponding subset of states. 
This construction allows approximating complex nonlinear dynamics by 
activating different communities at different levels  to model a particular type of linear dynamics at any given time, and using %
 smooth transitions between overlapping communities %
to model the smooth transitions between distinct linear dynamics.

To induce the structure of overlapping communities into $\Zmat$, the adjacency matrix of a directed random graph, and allow both the number of communities and number of nodes (dimension of $\Zmat$) to grow without bound, 
we propose the graph gamma process (GGP).
A draw from the GGP consists of countably infinite latent communities, each of which is associated with a positive weight indicating the overall activation strength of the community. These communities all share the same set of countably infinite nodes (states) but place different weights on how strongly a node is associated with a community. We describe the detail in what follows.

\subsection{Graph Gamma Process}\label{HGP}

Denote %
$Z(i,:)$ and $Z(:,i)$ as  row $i$ and  column $i$ of $\Zmat$, respectively. Since $\E[\xv_{t}\given \xv_{t-1}, \Wmat, \Zmat] = (\Wmat\odot \Zmat) \xv_{t-1}$, we have 
\ba{&\E[x_{ti}\given \xv_{t-1}, \Wmat, \Zmat] = %
(W(i,:)\odot Z(i,:))\xv_{t-1},\label{eq:Zrow}\\
&\E[\xv_{t+1}\given \xv_{t}, \Wmat, \Zmat] =\textstyle (W(:,i)\odot Z(:,i))x_{ti} +\sum_{j\neq i}^\infty (W(:,j)\odot Z(:,j))x_{tj},\label{eq:Zcol}
}
 which means $x_{ti}$ will be dependent on $\xv_{t-1}$ if ${Z(i,:)}$ contains non-zero elements, and it will influence $\xv_{t+1}$ if $Z(:,i)$ contains non-zero elements. To construct a nonparametric Bayesian model that removes the need to tune the hidden state dimension, our first goal is to allow $\Zmat$ to have an unbounded number of rows and columns, which  means that the model can support %
 countably infinite %
 state-specific factors $\dv_i$, with which  the mean of $\yv_t$ given $\xv_t$ is factorized as $\E[\yv_t\given \xv_t,\Dmat]=\Dmat\xv_t=\sum_{t=1}^\infty \dv_i x_{ti}$. 
 
 To achieve this goal, with  $c_\rho>0$ and $G_{0,\rho}$ defined as a finite and continuous base measure over a complete and separable metric space $\Omega$,
we first introduce a gamma process $G_{\rho}\sim\Gamma\mbox{P}(c_\rho,G_{0,\rho})$ on the product space $\mathbb{R}^+\times \Omega$, where $\mathbb{R}^+:=\{x:x>0\}$, such that 
for each subset $A\subset\Omega$, we have $G_\rho(A)\sim\mbox{Gamma}(G_{0,\rho}(A),1/c_\rho)$. 
The L\'evy measure of this gamma process can be expressed as $\nu(d\rho  d \dv ) = \rho^{-1}e^{-c_\rho\rho} d\rho G_{0,\rho}(d\dv)$.
A draw from this gamma process can be expressed as $G_\rho=\sum_{i=1}^\infty \rho_i d\dv_i$, consisting of countably infinite atoms (factors) $\dv_i$ %
with weights $\rho_i$.
We view $\dv_i$ as the factor loading vector for latent state $i$, and will make $\rho_i$ determine the number of nonzero elements  in $Z(i,:)$ and, consequently, how strongly $x_{ti}$, the activation of state $i$ at time $t$, is influenced by $\xv_{t-1}$ of the previous time. As the number of $\rho_i$ that are larger than an arbitrarily small constant $\epsilon$  follows a Poisson distribution with a finite mean as $\gamma_{0,\rho}\int_{\epsilon}^\infty \rho^{-1}e^{-c_\rho \rho} d\rho$, where $\gamma_{0,\rho}:=G_{0,\rho}(\Omega)$ is the mass parameter, this  can be used to express the idea that only a finite number of elements in $\{x_{ti}\}_{i=1,\infty}$ at time $t$ will be dependent on $\xv_{t-1}$ of the previous time. 

We further mark each $\rho_i$ with a degenerate gamma random variable, changing the L\'evy measure of the gamma process to that of a marked gamma process \citep{poissonp} as
$\nu(d\rho\, d \dv\, d\tau)  = \rho^{-1}e^{-c_\rho\rho} d\rho G_{0,\rho}(d\dv) \gamma_{0,\tau}\tau^{-1}e^{-c_{\tau}\tau}d\tau$; we express  a draw from this marked gamma process as $G_{\rho,\tau} =\sum_{i=1}^\infty {(\rho_i,\tau_i)\delta_{\dv_i}} $.
We will make $\tau_i$ determine the random number of nonzero elements in $Z(:,i)$ and, consequently,  how strongly $x_{ti}$, the factor score of state $i$ at time $t$, will influence $\xv_{t+1}$ of the next time point. As the number of $\tau_i$ that are larger than an arbitrarily small constant $\epsilon$  follows a Poisson distribution with a finite mean as $\gamma_{0,\tau}\int_{\epsilon}^\infty \tau^{-1}e^{-c_{\tau}\tau} d\tau$,
this can be used to express the idea that only a finite number elements in $\{x_{ti}\}_{i=1,\infty}$ at time $t$ will influence  $\xv_{t+1}$.

Given  $G_{\rho,\tau} =\sum_{i=1}^\infty {(\rho_i,\tau_i)\delta_{\dv_i}}$,
we further define  a gamma process $G_{o}\sim \Gamma\mbox{P}(c_o,G_{\rho,\tau})$,  with L\'evy measure $\nu(dr d\thetav d\psiv) = r^{-1}e^{-cr} dr G_o(d\thetav d\psiv)$,
a draw from which is expressed as
$G_o=\sum_{ \kappa =1}^\infty r_{\kappa}\delta_{\{\thetav_{\kappa}, \psiv_{\kappa}\}}$. In this random draw,  $r_{\kappa}\in\mathbb{R}_+$, reflecting the activation strength of community $\kappa$, is the weight of the $\kappa$th atom $\{\thetav_{\kappa}, \psiv_{\kappa}\}$, %
where $\thetav_{\kappa}=(\theta_{1\kappa},\ldots,\theta_{\infty \kappa})^T$, $\psiv_{\kappa}=(\psi_{1\kappa},\ldots,\psi_{\infty \kappa})^T$,  and $\theta_{i\kappa}$ and $\psi_{i\kappa}$,  representing how strongly that node  $i$ is associated with  community $\kappa$, are defined on $\rho_i$ and $\tau_i$, the weights of the atoms of the gamma process $G_{\rho,\tau}$, using
$$
\theta_{i\kappa}\sim\mbox{Gamma}(\rho_i,1/e),~~\psi_{i\kappa}\sim\mbox{Gamma}(\tau_i,1/f).
$$
 We refer to the hierarchical stochastic process constructed in this way as the GGP. We denote the mass parameter of the GGP as $\gamma_0:=\int G_o(d\thetav d\psiv)$. Inherited from the property of a gamma process, the GGP has an inherent shrinkage mechanism that its number of atoms (node communities) with weights greater than $\epsilon>0$ is a finite random number drawn from %
$\mbox{Pois}(\gamma_0\int_{\epsilon}^{\infty}{r^{-1}e^{-cr }dr}) $.

Given a random draw from the GGP as $G_o=\sum_{ \kappa =1}^\infty r_{\kappa}\delta_{\{\thetav_{\kappa}, \psiv_{\kappa}\}},$  we will let $r_{\kappa}$ determine the overall activation strength of community $\kappa$, $\theta_{i\kappa}$  how strongly state $i$ in community $\kappa$ is influenced by the states of the previous time in the same community, 
and $\psi_{j\kappa}$  how strongly state $j$ in community~$\kappa$ influences the states of the next time in the same community.  To express this idea, for  community~$\kappa$ parameterized by $\{r_{\kappa},\thetav_{\kappa},\psiv_{\kappa}\}$, we generate a community-specific sparse adjacency matrix, whose $(i,j)$th element is drawn as
\begin{align}
    &z_{ij\kappa}\sim\mbox{Bernoulli}(1-e^{-r_{\kappa}\theta_{i{\kappa}}\psi_{j{\kappa}}}). \label{eq:z_ijk} %
\end{align}
Thus from nodes $j$ to $i$, community $\kappa$ defines its own interaction probability, expressed as $p_{ij\kappa}=1-e^{-r_{\kappa}\theta_{i{\kappa}}\psi_{j{\kappa}}}$, and draws a binary edge $z_{ij\kappa}$  based on $p_{ij\kappa}$. While there are countably infinite nodes, in community $\kappa$, the total number of edges  is a finite random number and hence the number of nodes with nonzero degrees is also finite. 
\begin{lem}\label{lem:EZ}
The number of edges in community $\kappa$, expressed as $\sum_{i=1}^\infty \sum_{j=1}^\infty z_{ij\kappa} $, is finite. 
\end{lem}

As in \eqref{eq:z_ijk}, whether $z_{ij\kappa}=1$ or 0 is related to both the overall strength of community $\kappa$ and how strongly nodes $i$ and $j$ are affiliated with community $\kappa$. Lemma~\ref{lem:EZ}, whose proof is deferred to the Appendix, suggests that we can extract a finite submatrix $\mathcal{Z}_\kappa:=\{z_{ij\kappa}\}_{i,j\in \mathcal{S}_\kappa}$, where $\textstyle \mathcal{S}_\kappa:=\{i:\sum_j z_{ij\kappa}+\sum_j z_{ji\kappa}>0\}$ is the set of nodes with non-zero degrees in community $\kappa$. We consider $\mathcal{S}_\kappa$ as the nodes activated by community $\kappa$ and $\mathcal{Z}_\kappa$  as its nonempty graph adjacency matrix. Thus under the proposed GGP construction, different communities could overlap in the nodes belonging to their respective nonempty graph adjacency matrices, which means it is possible that $\mathcal{S}_\kappa \cap \mathcal{S}_{\kappa'} \neq \emptyset$ for $\kappa\neq \kappa'$. If $\mathcal{S}_\kappa \cap \mathcal{S}_{\kappa'}=\emptyset$, then we consider communities $\kappa$ and $\kappa'$ as two non-overlapping communities.

Our previous analysis tells that whether $z_{ij}=1$
determines not only whether state $i$ at a given time will be dependent of the states of the previous time, but also whether state $j$ at a given time will influence the states of the next time.
To express the idea that  whether $z_{ij}=1$ is collectively decided by all countably infinite communities, whose nonempty adjacency matrices could overlap in their selections of nodes, we 
take the OR operation over all elements in $\{z_{ij\kappa}\}_{\kappa}$ to define the adjacency matrix of the full model as
\beq
z_{ij}=\lor_{\kappa=1}^\infty z_{ij\kappa},\label{eq:OR}
\eeq
which means $z_{ij}=1$ if at least one $z_{ij\kappa}=1$, indicating community $\kappa$ places a directed edge from nodes $j$ to $i$,  and $z_{ij}=0$ otherwise. In a matrix format, we have $$\Zmat = \lor_{\kappa=1}^\infty \Zmat^{(\kappa)},$$ where $\Zmat^{(\kappa)}$ represents the graph adjacency matrix of community $\kappa$, whose $(i,j)$th element is $z_{ij\kappa}$.

We note that  marginalizing out $\{z_{ij\kappa}\}_{\kappa}$, we can directly draw  
the graph adjacency matrix defined in \eqref{eq:OR} as
\ba{
\Zmat\sim\mbox{Bernoulli}(1-e^{-\sum_{\kappa=1}^\infty r_{\kappa}\thetav_k\psiv_k^T}),\label{eq:BerPo}
}
which can also be equivalently generated under the Bernoulli-Poisson link \citep{epm_aistats2015} as $$\textstyle\Zmat=\delta(\Mmat\ge 1), ~\Mmat=\sum_{\kappa=1}^\infty \Mmat_\kappa,~\Mmat_\kappa\sim\mbox{Pois}( r_{\kappa} \thetav_{\kappa}\psiv_\kappa^T),$$ where $\delta(\cdotv)$  returns one if the condition is true and zero otherwise. While the graph defined by $\Zmat$ has countably infinite nodes,
the total number of edges is finite and hence the number of nodes with nonzero degrees is also finite; the proof of the following Lemma is deferred to the Appendix.

\begin{lem}\label{lem:EZ2}
The number of edges in $\Zmat$, expressed as $\sum_{i=1}^\infty\sum_{j=1}^\infty z_{ij}$, is finite.
\end{lem}
In summary, the GGP uses a gamma process
to support countably infinite node communities in the prior, and another marked gamma process to support countably infinite number of nodes (states) shared by these communities. The adjacency matrix of the generated random graph from the GGP can be either viewed as taking the OR operation over all community-specific binary adjacency matrices, or viewed as thresholding a latent count matrix that aggregates the activation strengths across all communities for each node pair. Under this model construction, with the inherent shrinkage mechanisms of the gamma processes, only a finite number of communities will contain edges between the nodes, and the nonempty communities overlap with each other on their selections of nonzero-degree nodes, the total number of which across all communities is finite.

\subsection{Overlapping Community based Model Interpretation and Visualization}\label{sec:vis}

As in \eqref{eq:BerPo}, the sparsity pattern $\Zmat$ of the latent state-transition matrix can be linked to a latent count matrix whose expectation is $\sum_{\kappa} r_{\kappa}\thetav_{\kappa}\psiv_{\kappa}^T$. Thus we can measure the activation strength of community $k$ relative to the combined activation strength of all communities using
\begin{equation}
  \mbox{\Amat}_{\kappa}:=  \frac{r_{\kappa}\thetav_{\kappa}  \psiv_{\kappa}^T}{\sum_{\kappa'=1}^{K}r_{\kappa'}\thetav_{\kappa'}  \psiv_{\kappa'}^T}.\label{eq:relativestrength}
\end{equation}
Due to the shrinkage property of the gamma process, only a small number of $r_{\kappa}$ are encourage to have non-negligible values. Therefore, there is often only a parsimonious set of active communities in the posterior.  
Since $\sum_{\kappa} \Amat_{\kappa} $ is a matrix whose elements are all equal to one, we can use the $\Amat_{\kappa}$'s to decompose the reconstruction of the time series under the proposed GGP-LDS into different sub-sequences, the $\kappa^{th}$ of which reveals the type of temporal patters captured by community $\kappa$. 

More specifically, given the model parameters and latent state representation $\{\xv_t\}_{t}$, taken from a posterior sample inferred by GGP-LDS, we define 
\ba{
\hat{\yv}_{t}^{(\kappa)} = \Dmat\hat{\xv}_{t}^{(\kappa)},~~\hat{\xv}_{t}^{(\kappa)} = [(\Wmat\odot \Zmat)\odot \Amat_\kappa]\xv_{t-1}. \label{eq:subseq}
}

Note we have $\E[\xv_t\given \Wmat,\Zmat,\xv_{t-1}] = \sum_{\kappa}\hat{\xv}_{t}^{(\kappa)}$ and $\E[\yv_t\given \Dmat,\Wmat,\Zmat,\xv_{t-1} ] = \sum_{\kappa} \hat{\yv}_{t}^{(\kappa)}$. Thus we can consider $\hat{\yv}_{1:T}^{(\kappa)}$  as a sub-sequence in the data space and $\hat{\xv}_{1:T}^{(\kappa)}$ as a sub-sequence in the latent space, both extracted according to the relative strength of community $\kappa$. Our experiments show that different dimensions of sub-sequence $ \hat{\yv}_{t}^{(\kappa)}$ often resemble each other in temporal patterns, but differ from each other in magnitudes, which is the reason why the aggregation of these sub-sequences, expressed as $\sum_{\kappa} \hat{\yv}_{t}^{(\kappa)}$, %
can exhibit distinct temporal behaviors at different data dimensions.  Note one may also define $\hat{\xv}_{t}^{(\kappa)} = (\Wmat\odot \Zmat^{(\kappa)})\xv_{t-1}$, which would lead to somewhat different community-specific sub-sequences, whose superposition no longer reconstructs $\E[\yv_t\given \Dmat,\Wmat,\Zmat,\xv_{t-1} ]$ but may help highlight transitional time points between different linear dynamics.

To visualize the overlapping community structure of the latent states inferred by the proposed GGP-(G)LDS, we order the communities in decreasing oder based on their overall activation strength, which, for example, can be measured by either $\|\Zmat_\kappa\|_0 = \sum_{i}\sum_j z_{ij\kappa}$ or $\|\Mmat_\kappa\|_1=\sum_{i}\sum_j m_{ij\kappa}$. In this paper, we rank the activation strength according to $\|\Mmat_\kappa\|_1$.
As in \eqref{eq:Zrow}, $Z(i,:)$ determines how latent state $i$ at the current time is influenced by the latent states of the previous time, and $Z(i,:)$ can be generated by thresholding latent counts $M(i,:)=\sum_{\kappa}M_\kappa(i,:)$. Thus we can we map row $i$ to community $$\pi_{row}(i)=\textstyle \arg\max_{\kappa} \sum_{j}m_{ij\kappa},$$
the primary community via which latent state $i$ is influenced by the latent states of %
 the previous time.
Similarly, as in \eqref{eq:Zcol}, $Z(:,j)$ determines how latent state $j$ at the current time is influencing the latent states of the next time, we can map column $j$ to community $$\pi_{col}(j)=\textstyle \arg\max_{\kappa} \sum_{i}m_{ij\kappa},$$
 the primarily community via which %
latent state $j$ influences the %
the latent states of the
 next time.
To help visualize $\Zmat$, we rearrange its rows such that the rows mapped to the same community according to $\pi_{row}(i)$ are placed in the same row block, and  we rearrange the columns in the same way according to $\pi_{col}(j)$.

For the rows inside the same row block ($i.e.$, sharing the same $\pi_{row}(i)$), we rearrange them in decreasing order based on $\sum_j m_{ij\pi_{row}(i)}$; we follow the same way to rearrange the columns inside each column block according to $\sum_i m_{ij\pi_{col}(j)}$.

Note that
$m_{ij\kappa}$ are augmented count variables in a posterior sample, %
which are introduced to provide closed-form Gibbs sampling update equations, as described below.

\subsection{Hierarchical Model and Bayesian Inference}\label{sec:STLDS}

To facilitate implementation, we truncate the GGP by setting $K$ as an upper-bound of the number of communities, and $S$ as an upper-bound of the number of states (nodes). We set $\gamma_{0,\rho}=\gamma_{0,\tau}=\gamma_{0}$. We  make the scales of $\theta_{i\kappa}$ and $\psi_{j\kappa}$ change with $\kappa$ to increase  model flexibility.
The hierarchical model of the truncated GGP-LDS is expressed as
\begin{align}
&\yv_{t}\sim \mathcal{N}(\Dmat  \xv_t,\,\Phimat^{-1}), ~ \quad \bold{\Phi}\sim \mbox{Wishart}(\bold{V},V+2),\notag\\
	&\xv_t\sim \mathcal{N}\left[(\bold{W} \odot \bold{Z}) \xv_{t-1},\,\mbox{diag}(\lambda_1,\ldots,\lambda_S)^{-1}\right],\notag\\
	& %
	\dv_{s}\sim \mathcal{N}\left(\mathbf{0},\,\bold{I}_{V}/\sqrt{V}\right), \quad  \lambda_s \sim\mbox{Gamma}\left(a,1/b\right) ,\notag\\
	& w_{ij}\sim \mathcal{N} (0 ,\varphi_{ij}^{-1}),\quad \varphi_{ij}\sim\mbox{Gamma}\left(\alpha_{0}, {1}/{\beta_{0}}\right),\notag\\
		& %
		z_{ij} = 
	\lor_{\kappa=1}^K z_{ij\kappa},\quad z_{ij\kappa}=\delta(m_{ij\kappa}\ge 1),\notag \\
	&m_{ij\kappa}\sim
    \mbox{Pois}\left(%
    r_{\kappa}\theta_{i\kappa}\psi_{j\kappa}\right),~
     \quad r_{\kappa}\sim \mbox{Gamma}({\gamma_{0}}/{K},{1}/{c}),\notag\\
	& \theta_{i\kappa}\sim\mbox{Gamma}\left(\rho_{i}, 1/e_{\kappa}\right), \quad \rho_{i}\sim\mbox{Gamma}\left(\gamma_{0}/S,1/c_{\rho}\right),\notag\\ &\psi_{j\kappa}\sim\mbox{Gamma}\left(\tau_{j},1/f_{\kappa}\right), \quad \tau_{j}\sim \mbox{Gamma}\left(\gamma_{0}/S,1/c_{\tau}\right),\label{eq:fullmodel}
\end{align} 
where 
$ %
e_{\kappa}, f_{\kappa}\sim\mbox{Gamma}\left(\alpha_0,1/\beta_0\right)$ and ${\xv_0}\sim \mathcal{N}(\mv_0,\Hmat_0)$. 
As in Lemma \ref{lem:EZ2}, the total number of nonzero elements in $\Zmat$ has a finite expectation. Thus if the GGP truncation levels $K$ and $S$ are set large enough, it is expected for some state $i$ that $\sum_{j}z_{ij}=0$, which means its corresponding row in $\Zmat$ has no nonzero elements, and/or $\sum_{j}z_{j i}=0$, which means its corresponding column in $\Zmat$ has no nonzero elements. 
If node $i$ has zero degree that $\sum_{j}z_{ij}=\sum_{j}z_{ji}=0$, then $x_{ti}$ will neither be dependent on $\xv_{t-1}$ nor influence $\xv_{t+1}$, which means  
 $\{x_{ti}\}_{t}$, the factor scores of state $i$,  capture only the non-dynamic noise component of the data. Moreover, the proposed model will penalize  the total energy captured by zero-degree node (state) $i$, expressed as $\sum_{t=1}^T x^2_{ti}$ if it is a zero-degree node (see Appendix \ref{proof_apx} for more details).
 
 To accommodate count data that are often overdispersed, we replace the Gaussian distribution based observation layer of GGP-LDS, $i.e.$, the first row of \eqref{eq:fullmodel}, with a negative binomial distribution based observation layer as 
 \begin{align}
	& \yv_t\sim\mbox{NB}  \left(\eta,\sigma \left( \Dmat \xv_t\right)\right), ~ \quad \eta \sim \mbox{Gamma} \left(\alpha_{\eta},1/\beta_{\eta}\right) , \quad \alpha_{\eta}, \beta_{\eta} \sim \mbox{Gamma} \left(\alpha_{0},1/\beta_{0}\right),\label{eq:nb_model}
\end{align}  
  while keeping the other parts of the model %
unchanged; we refer to the modified model for count observation as GGP generalized LDS (GGP-GLDS). Due to the use of the negative binomial distribution, we may also refer to the modified model as GGP negative binomial dynamical system (GGP-NBDS).

We perform Bayesian inference via Gibbs sampling. %
Exploiting a variety of data augmentation and marginalization techniques developed for discrete data \citep{zhou2013negative,epm_aistats2015} , we provide closed-form Gibbs sampling updated equations for all model parameters, as described in detail in  Appendix \ref{Gibbssampling}.  Unless specified otherwise, we consider 6000 Gibbs sampling iterations, treat the first 3000 samples as burnin, and collect one sample per 60 iterations afterwards, resulting in a collection of 50 posterior MCMC samples that are used to predict the means and uncertainty of future observations.  In the following section, we provide 
a review of related work, where we compare the proposed models against a variety of dynamical systems, including switching LDSs and autoregressive, nonparametric,  and deep neural network based models.

\section{Related Work} \label{related}
The main challenges for modeling time series include providing accurate forecast, having the ability to process missing data, and learning interpretable latent representation %
such that latent structures ($e.g.$, clusters) %
can be translated into meaningful  data temporal  patterns, representing different time series behaviors, 
without the need to perform data specific tuning. %
A time series method addressing all these challenges is desired to have %
the following properties:
1) The underlying model is neither  over- nor under-parameterized, which will result in over- or under-fitting, respectively, and hence poor performance in forecasting \citep{liu2015regularized,Kalantari2018nonparametric}. %
2) The model learns a parsimonious set of %
latent states such that the induced latent structures can explain the underlying patterns of the time series without the need of manual tuning.  %
Inducing too many (overlapping) clusters of the latent states makes the model difficult to interpret, while imposing heavy regularization can enhance interpretability but hinder prediction. A good model will be able to balance interpretability and predictive performance. 
3) The model is capable of %
modeling %
non-linear dynamics exhibited by the time series. %
4) The model is capable of capturing latent state transition behaviors and the uncertainty of %
 latent parameters.
In this section, we review related works %
and discuss whether they satisfy the aforementioned properties. %

Different from LDSs that employ latent state-transition blocks, autoregressive models that directly regress on the past observations are also commonly used to model real-valued time series data
\citep{harrison2003multivariate,davis2016sparse,saad2018temporally}, with a wide range of applications, such as health care and epidemic modeling \citep{li2010application,kennedy2011time}. In addition to LDSs and autoregressive models,  there also exist several other types of parametric models %
\citep{barber2011bayesian}. To achieve good performance on a given time series, these parametric models in general require searching over a large set of possible parameter settings or model configurations via cross validation, or appropriately  regularizing the training.

A variety of regularization techniques have been proposed for parameteric time series models. 
\citet{charles2011sparsity} introduce a sparsity constraint on the update equations of Kalman filtering to enforce the latent state vectors to be sparse, but need to assume that all the other model parameters, including the observation and transition matrices and noise covariance matrices, are known. \citet{stadler2013penalized} encourage a hidden Markov model (HMM) to have sparse state-specific inverse covariance matrices by imposing $L_{1}$-penalties on their elements.
\citet{liu2015regularized} propose to regularize the state transition matrix by penalizing its nuclear norm and imposing multivariate Laplacian priors over its rows. %
\citet{siddiqi2010reduced} propose a solution on reduced-rank HMM by relaxing the assumptions of a spectral learning algorithm by learning a $K$-dimensional subspace and finding the mapping between the high dimensional and low dimensional spaces.

Another set of time-series models are nonparametric Bayesian switching LDSs \citep{FoxSLDS09,pmlr-v54-linderman17a}, in which every temporal segment of the time series is fitted by one LDS. These models are focused on finding a mixture of LDSs, which are used to fit different time series segments, and a switching mechanism between different LDSs is learned to model the transitions between segments. Switching LDSs, however, may not provide satisfactory predictive performance on test data, as false switching, missed switching, and delayed switching could all compromise their predictions.
 \citet{chiuso2010learning} design another type of nonparametric Bayesian models that identify sparse linear systems. Unlike the proposed GGP-LDS, it  assumes no latent state transitions and models each observation as a linear combination of previous observations and some external input.  

In addition, there are
models that use the hierarchical Dirichlet process \citep{HDP} priors over the states in hidden Markov models \citep{johnson2013bayesian,FoxSLDS09}. There are also models that perform clustering on the time series use a Pitman-Yor process based mixture prior on non-linear state-space models  \citep{nieto2014bayesian}, and Dirichlet process mixtures \citep{caron2008noise} for modeling noise distributions. These models are not fully nonparametric as they typically have some  parametric assumptions as part of the model such as having a fixed number of  hidden states or imposing explicit specifications of the underlying temporal dynamics, such as seasonality and trends. %

\citet{chiuso2010learning} design a  nonparametric Bayesian  model  to identify sparse linear systems. %
It assumes no latent transitions and believes each observation is a linear combination of previous observations plus some external input. %
\citet{saad2018temporally}  introduce a recurrent Chinese restaurant process based mixture to capture temporal dependencies and a hierarchical prior to discover groups of time series whose underlying dynamics are modeled jointly. This model is able to cluster the  observations to a set of trajectories with similar behaviors, although it is prone to creating unnecessary clusters as if the same pattern  repeats with different magnitudes in two different segments of the observation, these two segments  are likely to be assigned to two different clusters. This may result in many unnecessary clusters for high dimensional and/or lengthy data.

Another widely used type of time series models are autoregressive models \citep{harrison2003multivariate,davis2016sparse,saad2018temporally}. %
There also exist several other parametric models, such as \citet{barber2011bayesian}, that %
provide additional tools to model time series. 
Most of these parametric models require searching over a large set of possible parameter settings with cross validation or model configurations to achieve satisfactory performance.

On the other side of the spectrum, we have a set of neural network based solutions, such as recurrent neural network based systems  \citep{flunkert2017deepar,lai2018modeling}, which are quite flexible but suffer from several major limitations. They often provide point estimate of their parameters, without uncertainty estimation, %
and often have the problem of interpretability and need considerable amount of data to reliably train the model.

In this paper, we introduce a nonparametric Bayesian hierarchical  model to address the aforementioned  challenges. The proposed model is based on the LDS and GGP. The proposed GGP is designed to learn overlapping communities of latent states such that each community models a behavior which can be described with a linear system. Instead of assigning one community (LDS) to one segment of the observed trajectory, our model allows multiple communities to be used at any time point. 
This allows the model to break the sophisticated behavior in a trajectory to a weighted combination of simpler behaviors modeled by different linear systems, and it helps to model the nonlinearities of the data using multiple linear systems and smooth transitions between them. %

\section{Experimental Results}

The code is provided at: {\color{blue} \url{https://github.com/GGPGLDS/GGP_GLDS}}.
In this section, we will demonstrate the interpretablity of GGP-(G)LDS and its predictive performance on several different datasets. Due to the inherent shrinkage mechanisms of the GGP, we find that the proposed nonparametric Bayesian model is not sensitive to the choice of the truncation levels $S$ and $K$ as long as they are set large enough.
For all the datasets in this section, we truncate them at $K = 16$ and $S = 30$, which are found to be large enough to accommodate all nonempty node communities, with 
interpretable latent representation and good predictive performance. 
Our Gibbs sampling based inference is not sensitive to initialization, allowing us to randomly initialize the model parameters. %
In this paper, we set $\gamma_0=\alpha_0=\beta_0=c=c_\rho=c_\tau=1$ for all experiments. We set $a=1$ and $b=0.1$ for all experiments (except for all visualizations, we set $b=1$ to encourage sparser latent state-transition matrices), %
encouraging $\lambda_s^{-1}$ to be small and hence encouraging the latent state representation vector to be constituted more by the autoregressive components and less by %
the white noise, generated by $\mathcal{N}\left[\bold{0},\mbox{diag}(\lambda_1,\ldots,\lambda_S)^{-1}\right]$. 

For GGP-(G)LDS, we measure the performance of $t$-step prediction, for  $t=1,\ldots, 10$. {In addition, 1-step-at-a-time predictive performance will be provided %
by using Kalman filter to  update the state $\xv_t$ after observing $\yv_t$, while other model parameters, including $\Dmat$, $\Phimat$, $\Wmat$, $\Zmat$, and $\lambdav$, stay the same to make the next step prediction. Note that 1-step-at-a-time prediction via Kalman filter is only an option readily available for an LDS based system (not including switching LDS).}
Denoting $\hat{\yv}_t$ as the $t$-step prediction of a given model, we 
 measure its $t$-step predictive performance using mean absolute error (MAE) for real time series as 
\ba{\mbox{MAE}_t=%
\sum _{v=1}^{V}\left|y_{dt}-\hat{y}_{dt}\right|/{V},\label{MAE}
}
  and use mean absolute relative error (MARE) for count time series as 
  \ba{\mbox{MARE}_t=%
\sum _{v=1}^{V}\frac{\left|y_{dt}-\hat{y}_{dt}\right|}{{V(y_{dt}+1)}}.\label{MRAE}
}
We compare the  predictive performance of GGP-GLDS with %
these of several representative time series models, whose description and
parameter setup for each dataset are described in detail below. 
For each dataset, we consider tuning important parameters for each competing algorithm. Notably for GGP-(G)LDS, 
when evaluating predictive performance, we simply use a same set of non-informative hyperparameters across all datasets and initialize all learnable parameters at random. 
Additional experiments on a synthetic dataset (the FitzHugh-Nagumo model) and a real dataset (closing stock price of twelve companies) will also be provided in the Appendix.

\subsection{Lorenz Attractor}\label{sec:Lorenz}
To demonstrate the performance of GGP-LDS on a dataset that has an underlying nonlinear dynamical pattern, we consider the Lorenz Attractor. We show how GGP-LDS finds an interpretable %
approximation to the generated time series with nonlinear dynamics. 
 The Lorenz system is a classical nonlinear differential equation with three independent variables, defined as $$
\frac{dx_1}{dt} = \alpha (x_2 -  x_1),~~\ \frac{dx_2}{dt} = x_1 (\beta - x_3) -x_2,~~ \frac{dx_3}{dt} = x_1 x_2 -  \gamma x_3 .$$ 
There exist %
approximate solutions for this differential equation \citep{hernandez2018novel,pmlr-v54-linderman17a,nassar2018tree}.
A linear approximation will be very useful as we can leverage for this non-linear system many canonical algorithms  developed for filtering and smoothing on linear systems. %
 To show how our model approximates the latent states, we generate numerical solutions of the Lorenz system with a randomly generated initial state, $\alpha = 1$, $\beta = 2$, $\gamma = 1$, and $T = 578$ time points. The original generated %
 variables under the Lorenz system 
 have three dimensions ($x_1, x_2$, and $x_3$). We treat them as latent variables and use a randomly generated $10\times3$ matrix to map them to  a 10-dimensional observation space. We use this $10 \times T$ observed data with added white Gaussian noise to train both GGP-LDS and a variety of baseline models.

\begin{figure}[h] 
  {\centering
  \begin{subfigure}[b]{1\linewidth}
 \begin{center}
 {%
 \includegraphics[width=100mm]{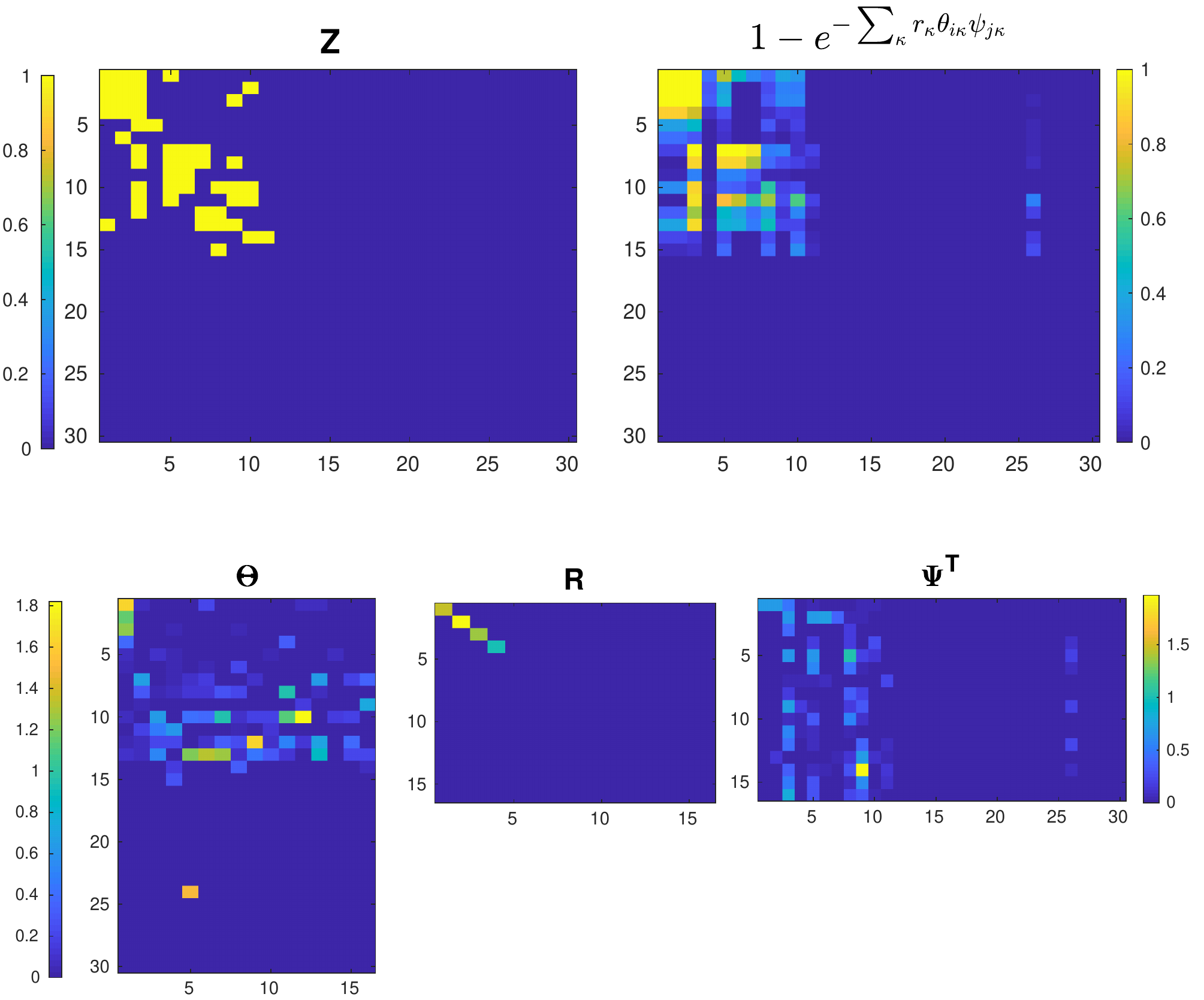}}\vspace{-1mm}%
 \caption{\small  %
}
  \label{fig:ZPThetaPsi}%
  \end{center}
  \end{subfigure}
  \\
  \begin{subfigure}[b]{1\linewidth}
  \begin{center}
{%
\includegraphics[width=110mm]{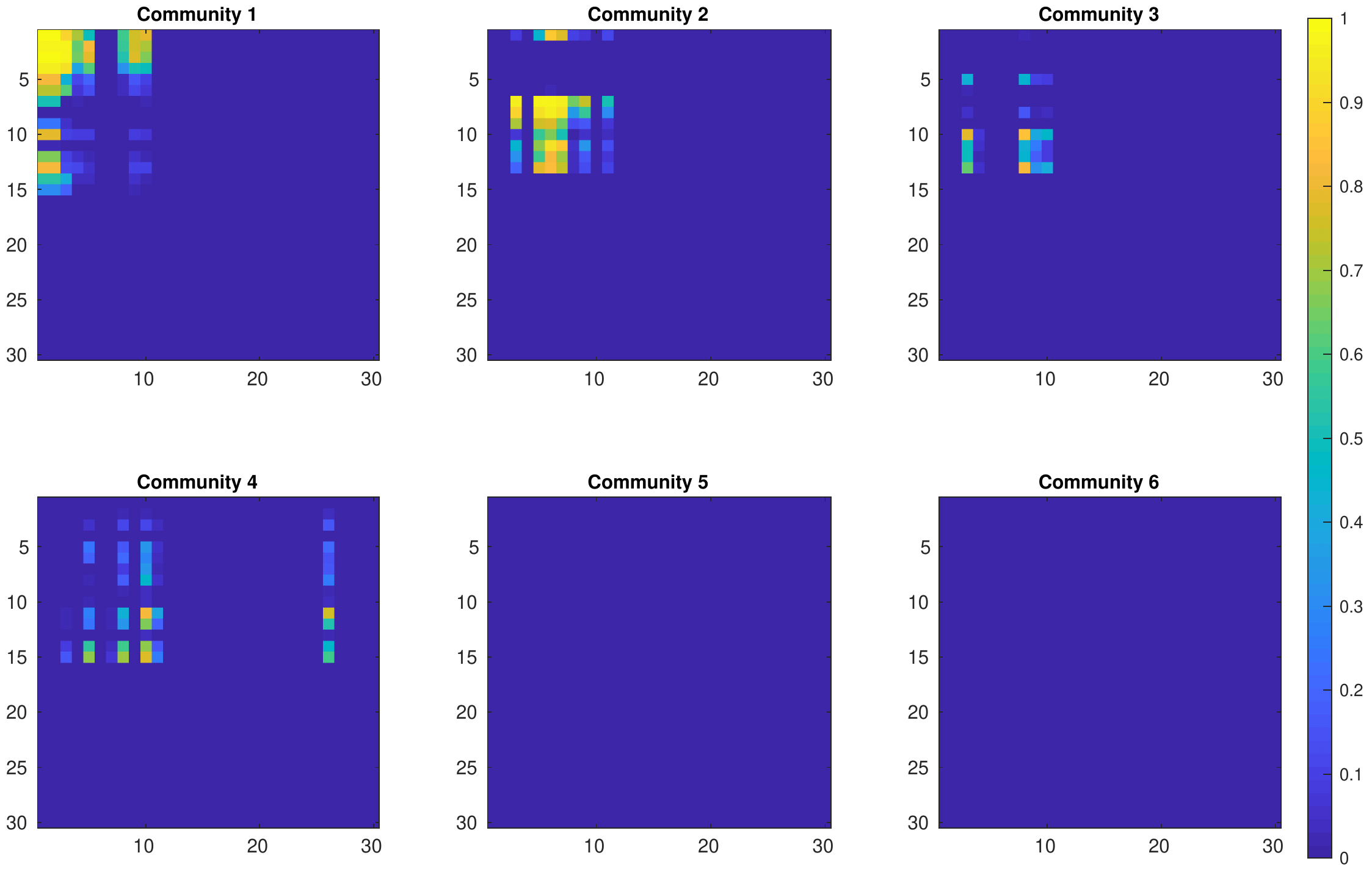}}%
\vspace{-1mm}%
\caption{\small } %
 \label{fig1_b}
 \end{center}
  \end{subfigure}
}\vspace{-6mm}\caption{\small \small (a): Visualization of a GGP-LDS inferred posteriori sample on a Lorenz Attractor synthesized time series. Top Left: $\Zmat$ from this posterior sample, where the rows and columns are separately reordered with the method described in Section \ref{sec:vis}; Top Right: The inferred activation probability of $\Zmat$; Bottom Left: $\Thetamat$, where $\theta_{i\kappa}$ %
shows how strongly state $i$  is influenced by the states of the previous time due to its association with  community $\kappa$;
Bottom Middle: $\Rmat$, whose diagonal elements show the activation strength of different  communities; Bottom Right: $\Psimat^T$, where $\psi_{\kappa,j}$ shows how strongly state~$j$  influences the states of the next time due to its association with community $\kappa$.\newline \indent\quad\, (b): Relative activation strength $\Amat_{\kappa}$, as defined in  \eqref{eq:relativestrength}, of the top six communities; note that $4$ active communities formed over $15$ active latent states are inferred by GGP-LDS while the truncation levels of the GGP are set as $K = 16$ and $S = 30$.
\label{fig1}
}
\end{figure}
\clearpage

Fig.~\ref{fig:ZPThetaPsi} illustrates a single posterior sample of GGP-LDS, focusing on the inferred graph adjacency matrix, %
 and the underlying activation probabilities of the edges of the graph adjacency matrix. %
 More specifically, 
 in the top row, we show on the left the graph adjacency matrix $\Zmat$, whose  rows and columns have been separately reordered following the description in Section \ref{sec:vis},  and on the right the underlying edge activation probabilities.
 In the bottom row, we show $\Thetamat=(\thetav_1,\ldots,\thetav_K)\in\mathbb{R}_+^{S\times K}$, %
 $\Rmat=\mbox{diag}(r_1,\ldots,r_K)\in\mathbb{R}_+^{K\times K}$, 
 and $\Psimat^T= (\psiv_1,\ldots,\psiv_K)^T\in\mathbb{R}_+^{K\times S}$,
 where $\theta_{i\kappa}$ shows the affiliation strength of $x_{(t+1)i}$ to the $\kappa^{th}$ community ($\kappa^{th}$ LDS) and $\psi_{\kappa, j}$ shows the association strength of $x_{tj}$ to the $\kappa^{th}$ LDS. 

 It can be observed how the shrinkage property of the gamma process $G_{\rho,\tau}$ has been effective in sparsifying the rows of  $\Thetamat$  and columns of $\Psimat^T$, with unnecessary elements being shrunk towards zero. In addition, it can be seen that each active row of $\Thetamat$, or active column of $\Psimat^T$ can potentially be a member of several different communities. The shrinkage property of the GGP $G_o$ drives many elements of $r_k$ towards zero and hence helps the model to pick which types of LDSs to be utilized. This is equivalent to say that the model infers which of these associations should be amplified or suppressed in expressing the underlying dynamics of the data. Moreover, %
 for the $\Thetamat$ matrix,  it has 7 members (rows) associated with community one, which implies there are 7 corresponding states at time $t+1$ that will be influenced by  $\xv_{t}$ of the previous time due to their associations with community one,  and $\Psimat^T$ shows that it has 4 members (columns) associated with community one, which implies that there are 4 corresponding states at time $t$ that will influence $\xv_{t+1}$ of the next time due to their associations with community one. Thus the transition matrix of the first member of overlapping LDSs will be the $7\times 4$ block shown on the top left corner, as shown in both $\Zmat$ and its corresponding probability matrix in~Fig.~1(a).

\begin{figure}[ht!] 
  \centering
  {\includegraphics[width=165mm]{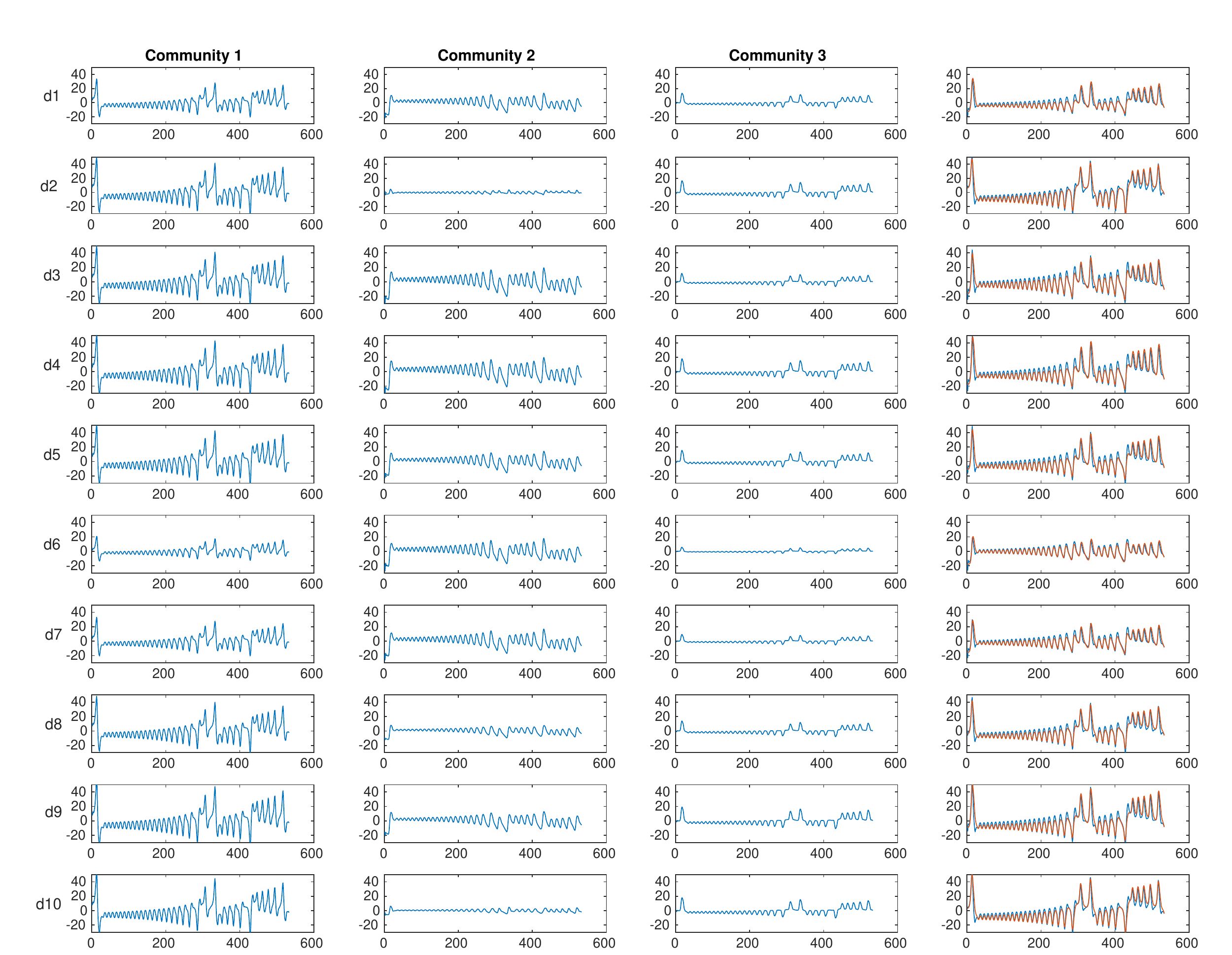}}%
\\
\caption{\small \small
Visualization of community specific sub-sequences decomposed from the Lorenz Attractor based time series reconstructed by GGP-LDS.
The first three columns show the sub-sequences of the three strongest communities, with each row showing one of the 10 data dimensions.
The last column shows the superposition of the three sub-sequences shown in the first three columns, where the observed time series is also included for comparison (highlighted in red). }
  \label{fig:lorenz2}
\end{figure}

\begin{figure}[t] 
\begin{subfigure}{.42\linewidth}
 \includegraphics[width=65mm]{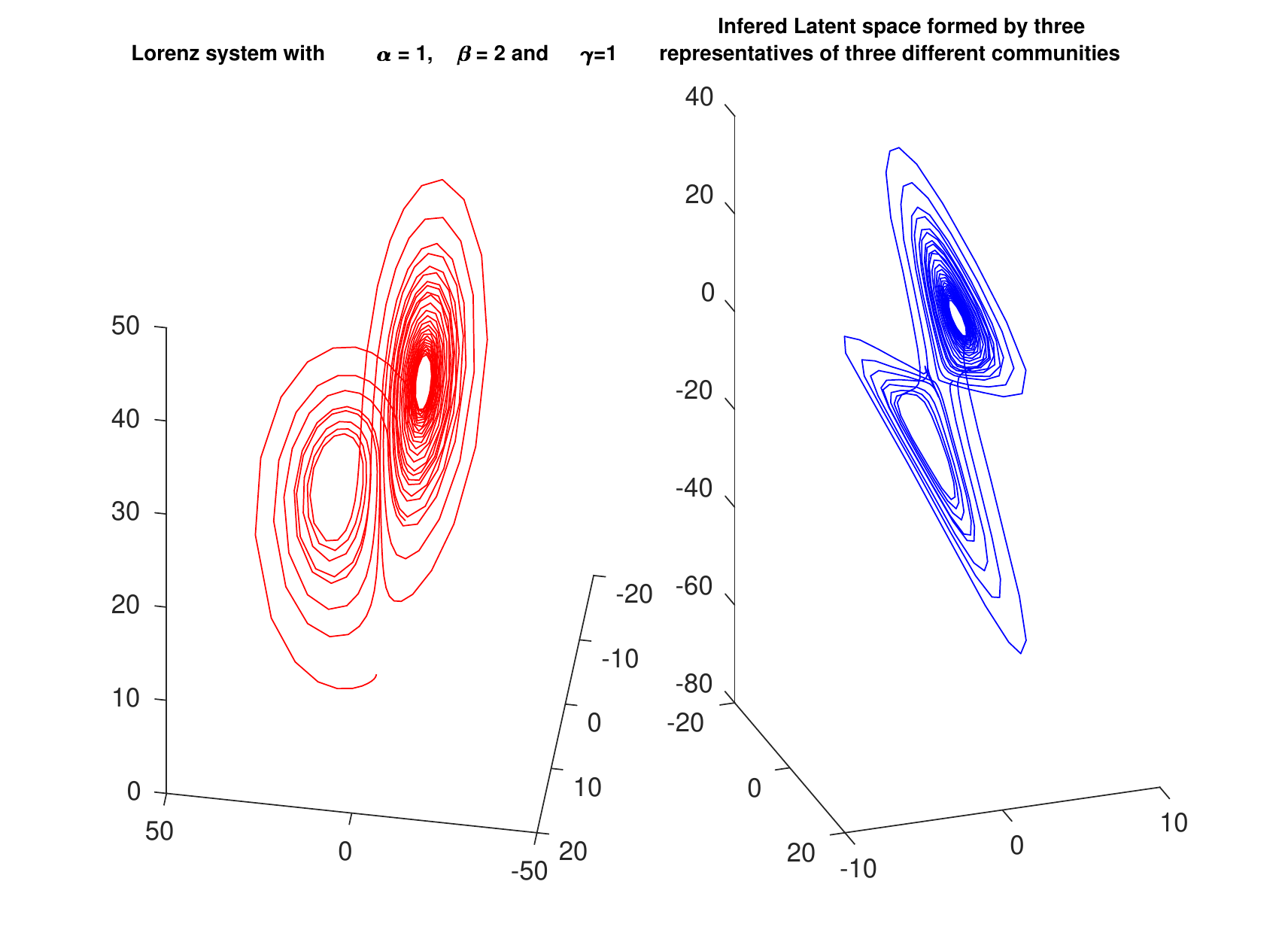}%
\caption{\small }
 \label{fig2:b}
  \end{subfigure} %
     \begin{subfigure}{.55\linewidth}
\includegraphics[width=95mm]{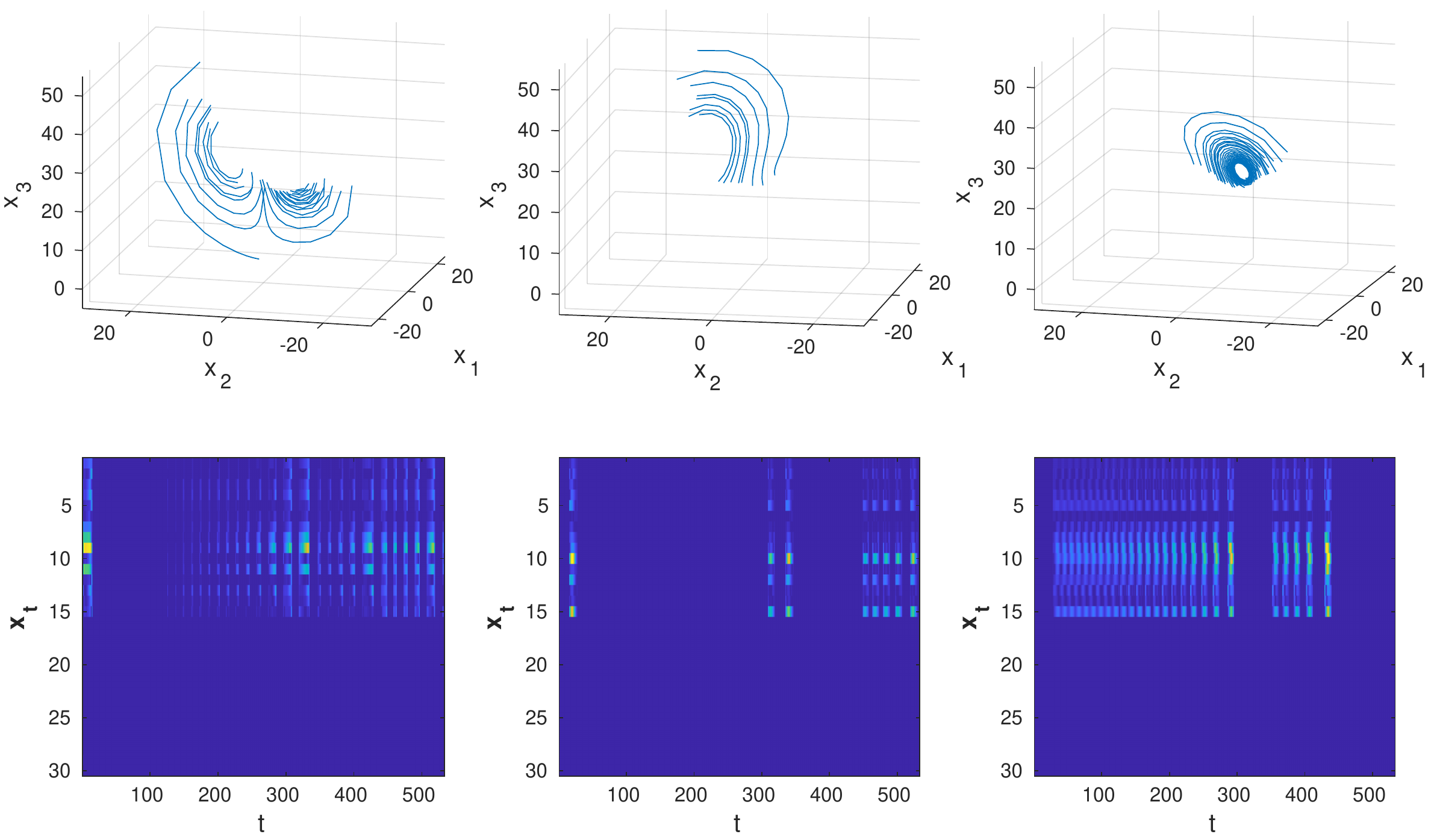}
   \caption{\small }
   \label{fig2_b}
  \end{subfigure}
  \caption{\small (a): The red trajectory shown on the left is synthesized by a Lorenz Attractor and used as the 3D latent state sequence to generate $\yv_{1:T}$, a 10D time series observation. Training GGP-LDS on $\yv_{1:T}$, the blue trajectory shown on the right is a 3D visualization of the inferred latent dynamics based on $(\hat{\xv}_{1:T}^{(1)}, \hat{\xv}_{1:T}^{(2)},\hat{\xv}_{1:T}^{(3)})$, the sub-sequences of the three strongest communities decomposed from the reconstructed time series by GGP-LDS.\\
\indent\quad\, (b): The bottom row visualizes the inferred latent states $\xv_{1:T}$ of GGP-LDS, which are  assigned into three non-overlapping clusters via the $K$-means algorithm, and  the top row  visualizes the corresponding segments of the  Lorenz Attractor synthesized 3D time series. }
   \label{fig:lorenz3}
\end{figure}

Out of $K=16$ (truncation level) possible  communities, we show in Fig.~\ref{fig:ZPThetaPsi_pedestrian} the top six formed communities, extracted from the inferred transition matrix for Lorenz Attractor, in six different subplots; %
we display each of these six communities using its relative strength defined in \eqref{eq:relativestrength}.
It is shown in Fig.~\ref{fig1_b} that our nonparametric Bayesian model finds four communities in total to model the underlying pattern of the data.  %
The number of linear solutions that our model has discovered is similar to that of \citet{nassar2018tree}, in which a tree based stick breaking process has been used as the prior. Moreover, it can be observed from Figs.~\ref{fig:ZPThetaPsi}  and \ref{fig1_b} that these 4 active communities are formed over 15 active states. Note for GGP-LDS, we have truncated its number of communities at $K=16$ and that of states at $S=30$. The results in Fig. \ref{fig1} have demonstrated the ability of GGP-LDS in inferring a parsimonious set of active communities and states to model the time series. 

Fig.~\ref{fig:lorenz2} %
shows how each community can reconstruct the observed data. %
Each row corresponds to a data dimension of the observed time series.
The first three columns show how the  three strongest communities contribute to data reconstruction, while the last column shows the  superpositions of the first three columns and compares them against the observed time series. It can be seen from Fig.~\ref{fig:lorenz2}  that the different dimensions of each community specific sub-sequence share similar temporal patterns, but may exhibit clearly different magnitudes.

In Fig.~\ref{fig2:b}, the red trajectory in the left plot represents %
the  Lorenz Attractor synthesized 3D time series that is used as the latent state representation to generate the observed 10D time series, and the  blue trajectory in the right plot illustrates a 3D representation of the latent dynamics of GGP-LDS trained on this 10D time series. More specifically, the blue trajectory is the visualization of the inferred community-specific latent sub-sequences ($\hat{\xv}_{1:T}^{(1)},\hat{\xv}_{1:T}^{(2)},\hat{\xv}_{1:T}^{(3)}$), where $\hat{\xv}_{1:T}^{(\kappa)}$, defined as in \eqref{eq:subseq}, is the latent sub-sequence extracted according to the relative strength of the $\kappa^{th}$ strongest community to the aggregation of all communities, as illustrated in Figs.~\ref{fig1_b} and \ref{fig:lorenz2} and described in detail in Algorithm~1.
It can be seen from Fig.~\ref{fig2:b} that the latent dynamics ($e.g.$, moving between two spirals) of GGP-LDS, visualized in 3D based on its inferred sub-sequences of its three strongest communities, are closely synchronized with the underlying dynamics of the Lorenz Attract synthesized 3D time series (a video showing how the red and blue trajectories move synchronously with each other has been included).
This shows that our model infers a close linear approximation to the underlying nonlinear dynamics. 
We provide another visualization of the latent dynamics inferred by GGP-LDS in Fig.~\ref{fig2_b}. Instead of decomposing the time series into sub-sequences, we now cluster it in time according to the inferred latent states $\xv_t$.
In the bottom row of Fig.~\ref{fig2_b}, the $\xv_t$'s are partitioned into three non-overlapping clusters with the $K$-means algorithm, which means each $\xv_t$ is assigned to one of the three clusters. In the top row of Fig.~\ref{fig2_b}, the same cluster assignment is applied to segment the Lorenz Attractor time series into three sequences that do not overlap in time. 
It is clear that the segmentation points based on the $\xv_t$'s inferred by GGP-LDS well align with the switching points between different linear dynamics, demonstrating the ability of GGP-LDS to seamlessly transit between different temporal patterns, each of which is modeled by adjusting the activation strengths of different latent state communities that can co-occur at the same time.

\begin{table*}[t]
  \caption{\small Lorenz Attractor predictive performance. The best result and the results that are not considered as statistically different %
  are highlighted in bold. }
  \label{Lorenz_prediction}
  \centering
  \vspace{-2mm}
  \resizebox{1.0\columnwidth}{!}{
  \begin{tabular}{lllllllllll}
    \toprule
    \multicolumn{11}{c}{Mean absolute error for 10 forecast horizons}                   \\
    \cmidrule(r){1-11}
    
    Algorithm & $t=1$ & $t=2$  & $t=3$  & $t=4$  & $t=5$   & $t=6$ & $t=7$  & $t=8$  & $t=9$  & $t=10$\\
    \midrule
    LDS 	 &$11.12_{(1.10)}$	&$13.76_{(2.47)}$	&$18.54_{(2.78)}$	&$22.34_{(2.90)}$	&$18.81_{(3.18)}$  &$13.12_{(3.30)}$	&$9.68_{(2.87)}$	&$8.54_{(2.37)}$	&$7.21_{(2.51)}$	&$7.54_{(2.48)}$    \\
    
    rLDSg     		  &$8.12_{(0.90)}$	&$12.76_{(1.31)}$	&$17.33_{(1.98)}$	&$20.52_{(1.61)}$	&$15.62_{(2.05)}$  & $10.54_{(3.21)}$ & $8.62_{(3.28)}$ & $9.42_{(3.65)}$ & $7.46_{(3.16)}$ & $6.28_{(3.49)}$   \\

    rLDSr     		  & $12.46_{(1.72)}$ & $19.22_{(3.72)}$ & $21.51_{(4.22)}$ & $25.32_{(3.67)}$ & $18.21_{(3.11)}$    & $11.51_{(4.16)}$ & $13.21_{(4.21)}$ & $10.21_{(4.11)}$ & $8.57_{(3.95)}$ & $6.91_{(3.18)}$  \\
    SGLDS    		  & $8.84_{(1.32)}$ & $10.43_{(1.66)}$ & $14.51_{(2.43)}$ & $15.32_{(3.61)}$ & $16.31_{(3.24)}$    & $15.32_{(3.83)}$ & $12.21_{(3.94)}$ & $9.13_{(4.24)}$ & $9.57_{(3.95)}$ & $7.91_{(3.68)}$  \\
    
    TrSLDS     		  & $5.21_{(0.62)}$ & $5.76_{(0.98)}$ & $6.23_{(1.31)}$ & $7.45_{(1.42)}$ & $\mathbf{5.31_{(1.19)}}$    & $\mathbf{5.12_{(1.53)}}$ & $\mathbf{4.21_{(1.24)}}$ & $\mathbf{2.31_{(1.18)}}$ & $\mathbf{2.57_{(1.45)}}$ & $\mathbf{5.78_{(1.11)}}$  \\

    Multi-output GP 	  & $11.52_{(1.58)}$ & $15.35_{(1.73)}$ & $16.21_{(2.21)}$ & $19.08_{(2.36)}$ & $17.21_{(3.79)}$  & $12.37_{(3.77)}$ & $8.38_{(3.98)}$ & $8.21_{(2.98)}$ & $6.31_{(3.21)}$ & $7.21_{(3.98)}$ \\
    
    FB Prophet     & $5.57_{(1.31)}$ & $11.82_{(1.45)}$ & $13.42_{(1.98)}$ & $15.21_{(2.04)}$ & $16.26_{(1.76)}$  & $9.41_{(1.86)}$ & $8.78_{(2.01)}$ & $7.66_{(1.91)}$ & $6.54_{(2.14)}$ & $6.72_{(2.23)}$  \\
    
    DeepAR 	  & $9.42_{(0.26)}$ & $10.21_{(0.31)}$ & $16.22_{(0.54)}$ & $16.42_{(1.28)}$ & $15.24_{(1.21)}$  & $11.21_{(1.61)}$ & $13.25_{(2.08)}$ & $12.83_{(2.83)}$ & $14.21_{(3.01)}$ & $16.25_{(3.21)}$\\
    
    TRCRP     & $5.66_{(0.86)}$ & $7.91_{(1.01)}$ & $11.23_{(1.35)}$ & $15.37_{(2.31)}$ & $16.21_{(2.42)}$   & $9.68_{(2.68)}$ & $8.21_{(2.98)}$ & $6.85_{(2.71)}$ & $5.63_{(2.38)}$ & $7.35_{(2.81)}$ \\
    
    GGP-LDS (10 steps)    &$\mathbf{ 2.12_{(0.84)}}$	&$\mathbf{3.76_{(1.87)}}$	&$\mathbf{4.77_{(2.68)}}$	&$\mathbf{5.04_{(3.16)}}$	&$\mathbf{4.83_{(3.32)}}$  &$\mathbf{4.50_{(3.27)}}$	&$\mathbf{4.15_{(3.34)}}$	&${4.14_{(3.51)}}$	&${4.60_{(3.66)}}$	&$\mathbf{5.24_{(3.86)}}$\\
    
  \cmidrule(r){2-11}  
    GGP-LDS (1 step)	&$2.10_{(0.52)}$	&$0.37_{(0.23)}$	&$0.32_{(0.17)}$	&$0.41_{(0.18)}$	&$0.40_{(0.24)}$ &$0.64_{(0.27)}$	&$0.84_{(0.28)}$	&$0.81_{(0.25)}$	&$0.66_{(0.23)}$	&$0.57_{(0.23)}$\\
    \bottomrule
  \end{tabular}}\label{tbl:lorenz_main}
\end{table*}

\begin{figure}[t] 
\vspace{0mm}
  \centering
\hspace{-8mm}%
\vspace{-0mm}
  {\includegraphics[width=144mm]{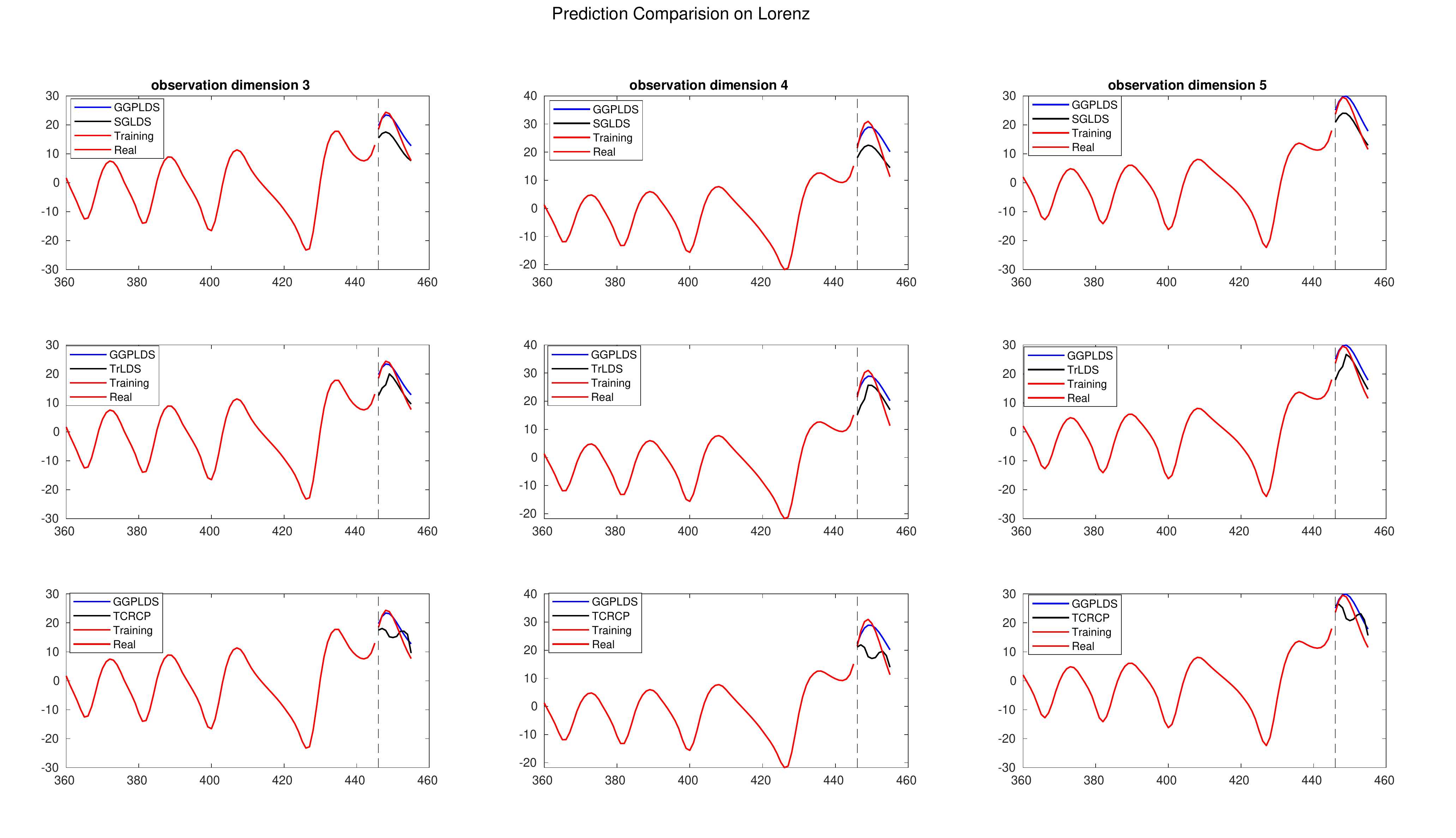}}
  
\caption{\small Comparison of predictive performance of the GGPLDS to SGLDS, TrSLDS and TRCRP on Lorenz attractor data. Each column of the figure represents one dimension of the observation (out of 10) and each row contain the visual comparison of GGP-LDS with one of those aforementioned algorithms.}
  \label{fig:Comparsion}
\end{figure}

In addition to these qualitative analyses, we quantitatively compare GGP-LDS and a variety of baseline algorithms on their predictive performance on the same 10D time series $\yv_t$, generated by adding Gaussian noise to $\Dmat \xv_{t}$, where $\xv_{1:T}$ is a Lorenz Attractor synthesized 3D time series. 
Below we list the parameter settings of the baseline algorithms; for each competing algorithm, we choose its setting on each dataset that provides it the best overall prediction performance over 10 forecast horizons, except for SGLDS and GGP-LDS that are insensitive to hyperparameter settings under their respective nonparametric Bayesian constructions:  %
\begin{itemize}
\item LDS of \citet{ghahramani96parameterestimation}, inferred by EM, with $K=6$.
\item $rLDS_g$ and $rLDS_r$  of \citet{liu2015regularized}, with $K=8$ and random initialization.
\item SGLDS of \citet{Kalantari2018nonparametric}, with the number of hidden states truncated at $K=30$.
\item TrSLDS of \citet{nassar2018tree}, with its tree depth set as 2 and dimension of latent states for each LDS as 4.
\item Multi-outputGP of \citet{alvarez2009sparse}, setting:
1. multi-gp option,
2. kerneltype: ggwhite,
3. optimizer: scg,
4. nlf = 1.
\item FB Prophet  of \citet{taylor2018forecasting},  default setting.
\item Deep AR of \citet{flunkert2017deepar},  default setting.
\item TRCRP of \citet{saad2018temporally}, with the Markov chain order set as $p=10$ .
\end{itemize}

As $t=445$ is one of the switching time points at which $\xv_t$ moves from one spiral to another, we choose $\yv_{1:445}$ for training. This set up can measure how well an algorithm detects and responds to changes in the underlying dynamics. 
The predictive performance of each algorithm is measured by mean absolute error defined in \ref{MAE} over a horizon of 10 time points. The results are presented in Table~\ref{tbl:lorenz_main}. 
As shown in Table \ref{tbl:lorenz_main},
most of the competing algorithms are not making good predictions following the switching point, likely because they expect that the trajectory will keep following the same spiral pattern before the switching point. In reality, the trajectory quickly switches to the other spiral pattern for a few steps before coming back to the same spiral pattern observed before $t=445$.
To further illustrate this point, 
we pick three different dimensions of the 10D time series $\yv_{t}$, and show in Fig.~\ref{fig:Comparsion} the prediction of four different algorithms, including SGLDS, TrLDS, TCRCP, and the proposed GGP-LDS,  on these three dimensions over a horizon of 10 time points.
It is evident from Fig.~\ref{fig:Comparsion} that at the switching point, SGLDS, TrLDS, and TCRCP all fail to detect the transition from one spiral to another. More specifically, 
SGLDS closely follows the pattern of the same spiral, TrSLDS is experiencing delays in switching to the correct LDS that better fits the second  spiral, and TRCRP creates wrong patterns. 

\subsection{FitzHugh-Nagumo}
\begin{figure}[t!] 
  \centering
 \subcaptionbox{\label{fig8:c}}
 {\includegraphics[width=80mm]{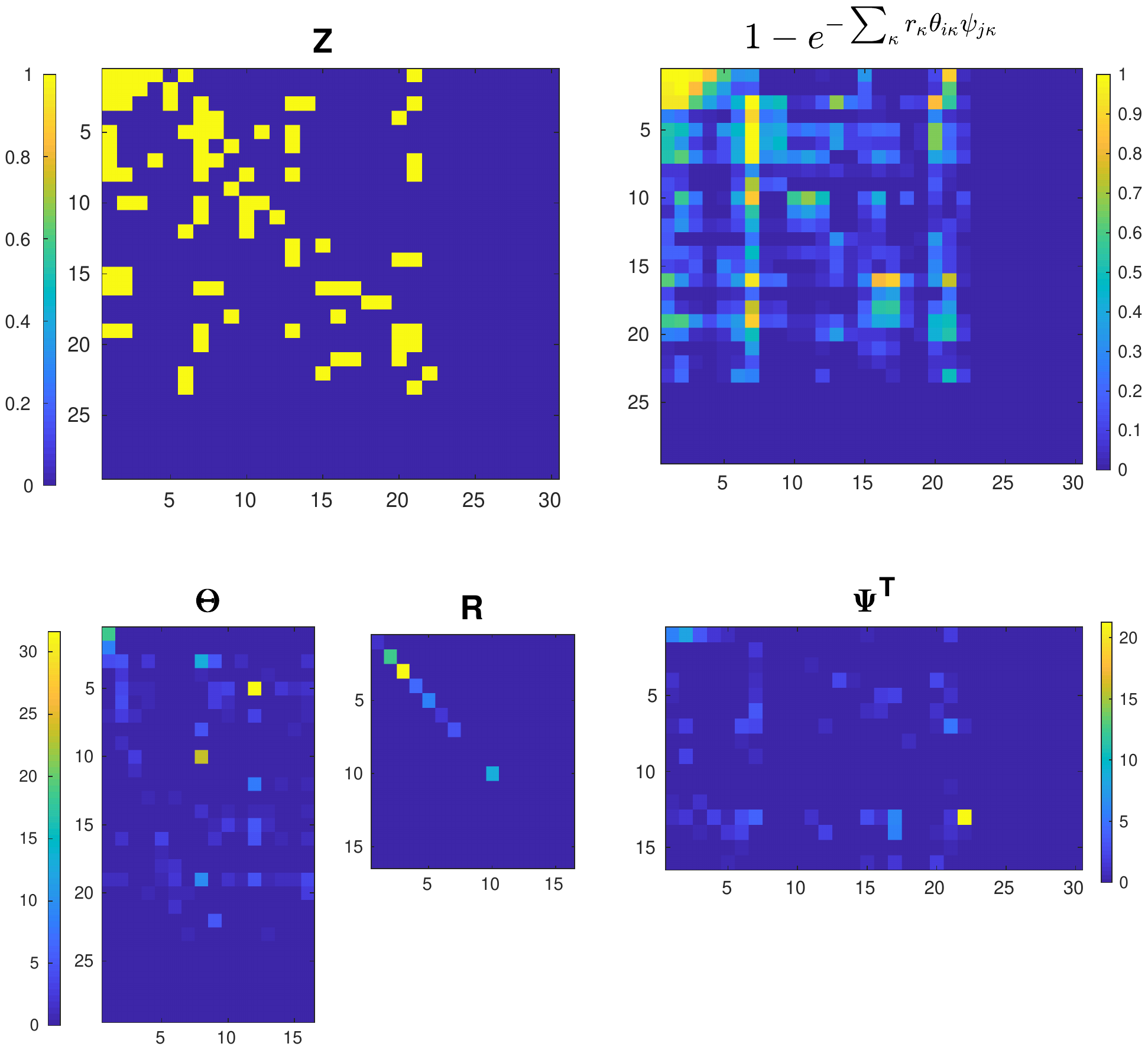}}
\quad\,\,\,\,%
  \subcaptionbox{\label{fig8:a}}
 {\, \includegraphics[width=74mm]{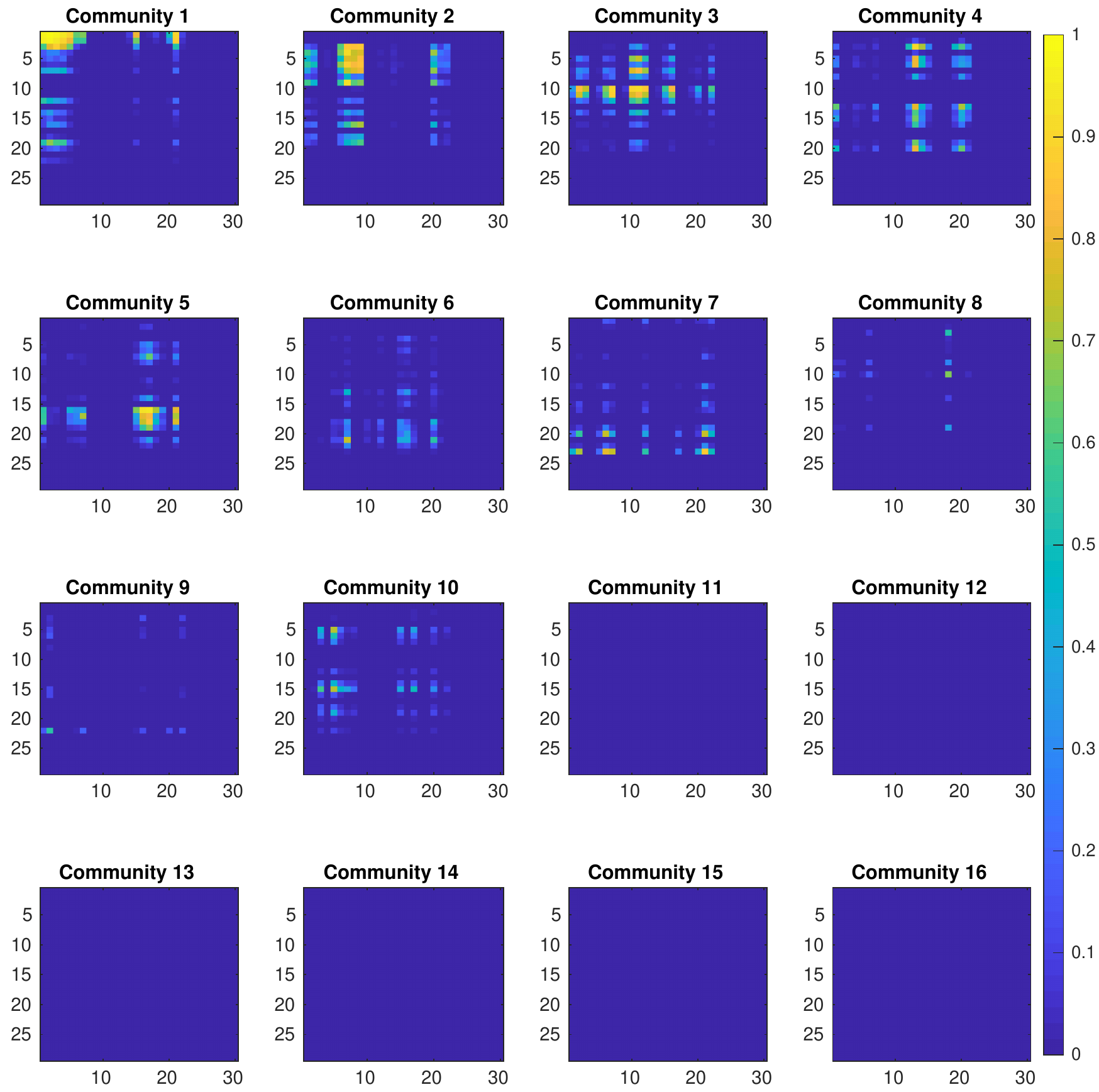}}\\\vspace{4mm}
 \subcaptionbox{\label{fig8:b}}{\includegraphics[width=80mm]{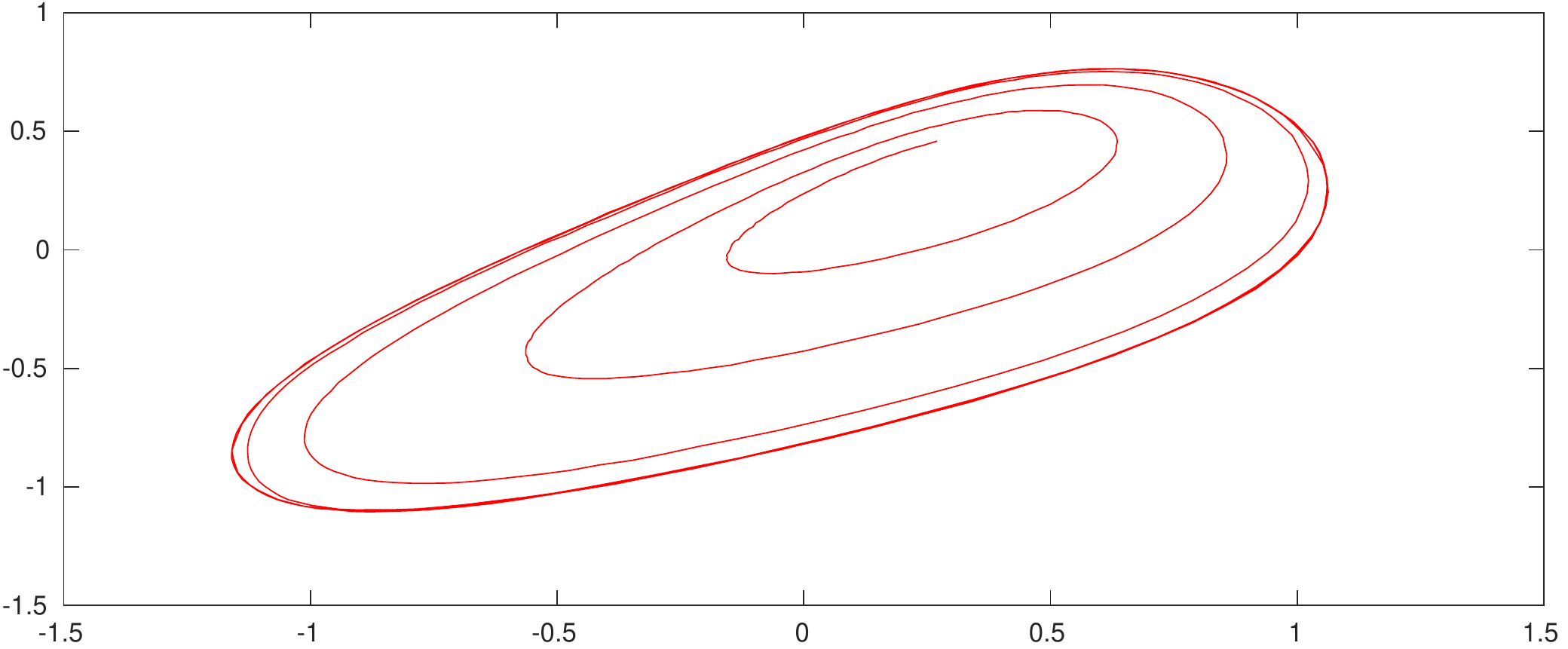}\quad
 \includegraphics[width=80mm]{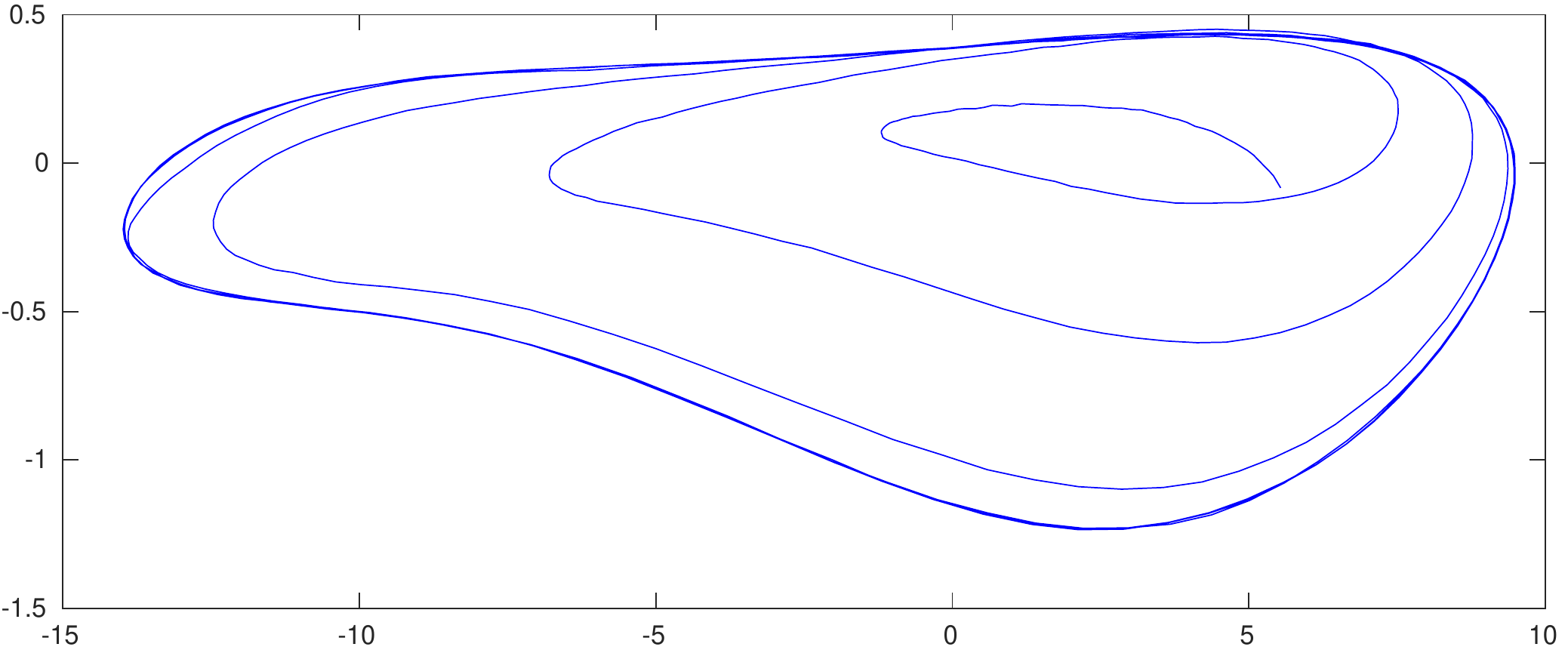}
 }%
\caption{\small Visualization of a posterior sample of GGP-LDS applied to FitzHugh-Nagumo, where (a), (b), and (c) are analogous plots to Figs. \ref{fig:ZPThetaPsi}, \ref{fig1_b}, and \ref{fig2:b}, respectively.
  \label{fig:FHN1}} %
\end{figure}

We provide another set of visualization of GGP-LDS by considering 
the FitzHugh-Nagumo (FHN) model, %
which can be formulated by the following differential equations:
\begin{align}
& \frac{dv}{dt}=v - \frac{v^3}{3} - w+ I ,\notag\\
&  \frac{dw}{dt}=c\times \left(v+a-bw\right).\notag
\end{align} 
 We set $I= 0.3$, $a= 0.7$, $b=0.8$, and $c=0.7$. We train our model with a trajectory whose starting point is picked randomly. The trajectory consists of 800 time points. The observation model is linear and Gaussian where $\dv_i\sim \mathcal{N} \left([0,0]^T,\left(\begin{matrix}1 & 0\\ 0 & 1\end{matrix}\right)\right)$ and $\Phimat= 0.01*\mathbf{I}_2$, where $\mathbf{I}_2$ is an identity matrix of dimension $2\times 2$. We provide a set of plots analogous to those in Figs \ref{fig:ZPThetaPsi}, \ref{fig1_b}, and \ref{fig2:b} for Lorenz Attractor. It can be seen in Fig. \ref{fig:FHN1} how the latent trajectories have been recovered by %
 ($\hat{\xv}_{c_1},\hat{\xv}_{c_2}$) that represent the activation strength of the two strongest communities $(c_1,c_2)$.

\subsection{Pedestrians' Trajectories}
We analyze a dataset that records by camera the 3D motions of pedestrians and their interactions, downloaded from \url{https://motchallenge.net/} and available in the provided code repository.  %
For visualization purpose, we use only 2 dimensional data for each pedestrian $(x,y)$. We select six pedestrians over 120 time points to train our model. The next 10 time points are used to measure the predictive performance of various algorithms.

Table \ref{Pedestrian}  compares the predictive performance of various algorithms  on this dataset. In most of the 10 forecast horizons our model has outperformed the other competing models. The parameter settings of various algorithms are the same as those in Section \ref{sec:Lorenz} for Lorenz Attractor, with the following exceptions:
\begin{itemize}
\item LDS of \citet{ghahramani96parameterestimation}, inferred by EM, with $K=8$.
\item TrSLDS of \citet{nassar2018tree}, with its tree depth set as 2 and dimension of latent states for each LDS as 2.
\end{itemize}

  \begin{table*}[t]
  \caption{\small Analogous table to Table \ref{Lorenz_prediction} for the Pedestrians dataset}
  \label{Pedestrian}
  \centering
  \resizebox{1\columnwidth}{!}{
  \begin{tabular}{lllllllllll}
    \toprule
    \multicolumn{11}{c}{Mean absolute error for 10 forecast horizons}                   \\
    \cmidrule(r){1-11}
    
    Algorithm & $t=1$ & $t=2$  & $t=3$  & $t=4$  & $t=5$   & $t=6$ & $t=7$  & $t=8$  & $t=9$  & $t=10$\\
    \midrule
    LDS 	   & $0.21_{(0.09)}$ 	&$0.31_{(0.17)}$	 	&$1.10_{(0.41)}$		&$1.27_{(0.46)}$	&$1.55_{(0.66)}$ & $1.48_{(0.65)}$ & $1.52_{(0.62)}$ & $1.62_{(0.61)}$ & $1.60_{(0.59)}$ & $1.76_{(0.71)}$  \\ 
    
    rLDSg     		  & $0.19_{(0.07)}$ 	&$0.29_{(0.12)}$	 	&$0.41_{(0.19)}$	&$0.71_{(0.16)}$	&$1.11_{(0.23)}$  & $1.08_{(0.59)}$ & $1.12_{(0.65)}$ & $1.27_{(0.58)}$ & $1.37_{(0.61)}$ & $1.41_{(0.70)}$   \\

    SGLDS    		  & $0.17_{(0.12)}$ & $0.24_{(0.16)}$ & $0.28_{(0.23)}$ & $0.42_{(0.21)}$ & $0.54_{(0.34)}$    & $0.66_{(0.31)}$ & $0.71_{(0.41)}$ & $0.87_{(0.44)}$ & $0.97_{(0.55)}$ & $1.02_{(0.68)}$  \\
    
    TrSLDS     		& $0.16_{(0.10)}$ & $0.26_{(0.12)}$ & $0.32_{(0.18)}$ & $0.40_{(0.21)}$ & $0.52_{(0.28)}$    & $0.60_{(0.29)}$ & $0.69_{(0.32)}$ & $0.79_{(0.39)}$ & $0.79_{(0.41)}$ & $0.92_{(0.48)}$  \\

    Multi-output GP 	  & $0.22_{(0.11)}$ & $0.28_{(0.16)}$ & $0.42_{(0.21)}$ & $0.60_{(0.20)}$ & $0.68_{(0.21)}$  & $0.89_{(0.17)}$ & $1.03_{(0.23)}$ & $1.15_{(0.28)}$ & $1.41_{(0.32)}$ & $1.40_{(0.30)}$ \\
    
    FB Prophet    & $0.17_{(0.05)}$	 &$0.26_{(0.08)}$	 	&$0.33_{(0.11)}$	&$0.40_{(0.12)}$	&$0.56_{(0.14)}$   & $0.65_{(0.14)}$ & $0.87_{(0.13)}$ & $0.86_{(0.17)}$ & $1.04_{(0.23)}$ & $1.12_{(0.20)}$  \\
    
      DeepAR 	  & $0.22_{(0.06)}$ & $0.21_{(0.04)}$ & $0.32_{(0.04)}$ & $0.42_{(0.12)}$ & $0.64_{(0.21)}$  & $0.81_{(0.27)}$ & $0.85_{(0.31)}$ & $0.98_{(0.36)}$ & $1.03_{(0.45)}$ & $1.15_{(0.51)}$\\
    
    TRCRP     & $0.17_{(0.09)}$ 	&$0.19_{(0.10)}$	 	&$0.29_{(0.12)}$		&$\mathbf{0.27_{(0.16)}}$	&$\mathbf{0.33_{(0.21)}}$    & $\mathbf{0.38_{(0.15)}}$ & $0.61_{(0.22)}$ & $0.66_{(0.28)}$ & $0.71_{(0.32)}$ & $0.92_{(0.38)}$    \\
    
    GGP-LDS (10 steps)     & $\mathbf{0.14_{(0.11)}}$ &$\mathbf{0.20_{(0.16)}}$	 	&$\mathbf{0.26_{(0.21)}}$	&$0.31_{(0.26)}$	&$0.38_{(0.32)}$   & $0.44_{(0.37)}$		&$\mathbf{0.52_{(0.44)}}$		&$\mathbf{0.59_{(0.50)}}$ 	&$\mathbf{0.67_{(0.57)}}$		&$\mathbf{0.75_{(0.64)}}$ \\
    
  \cmidrule(r){2-11}  
    GGP-LDS (1 step)	& $0.11_{(0.08)}$ &$0.13_{(0.09)}$	 	&$0.14_{(0.11)}$	&$0.14_{(0.11)}$	&$0.15_{(0.12)}$  &$0.17_{(0.13)}$ &$0.19_{(0.14)}$	&$0.21_{(0.16)}$	&$0.23_{(0.18)}$	&$0.26_{(0.21)}$ \\
    \bottomrule
  \end{tabular}}\label{tbl:pedestrian}
\end{table*}

\begin{figure}[t]
  \begin{subfigure}[b]{.5\linewidth}
   \begin{center}
 {\includegraphics[width=78mm]{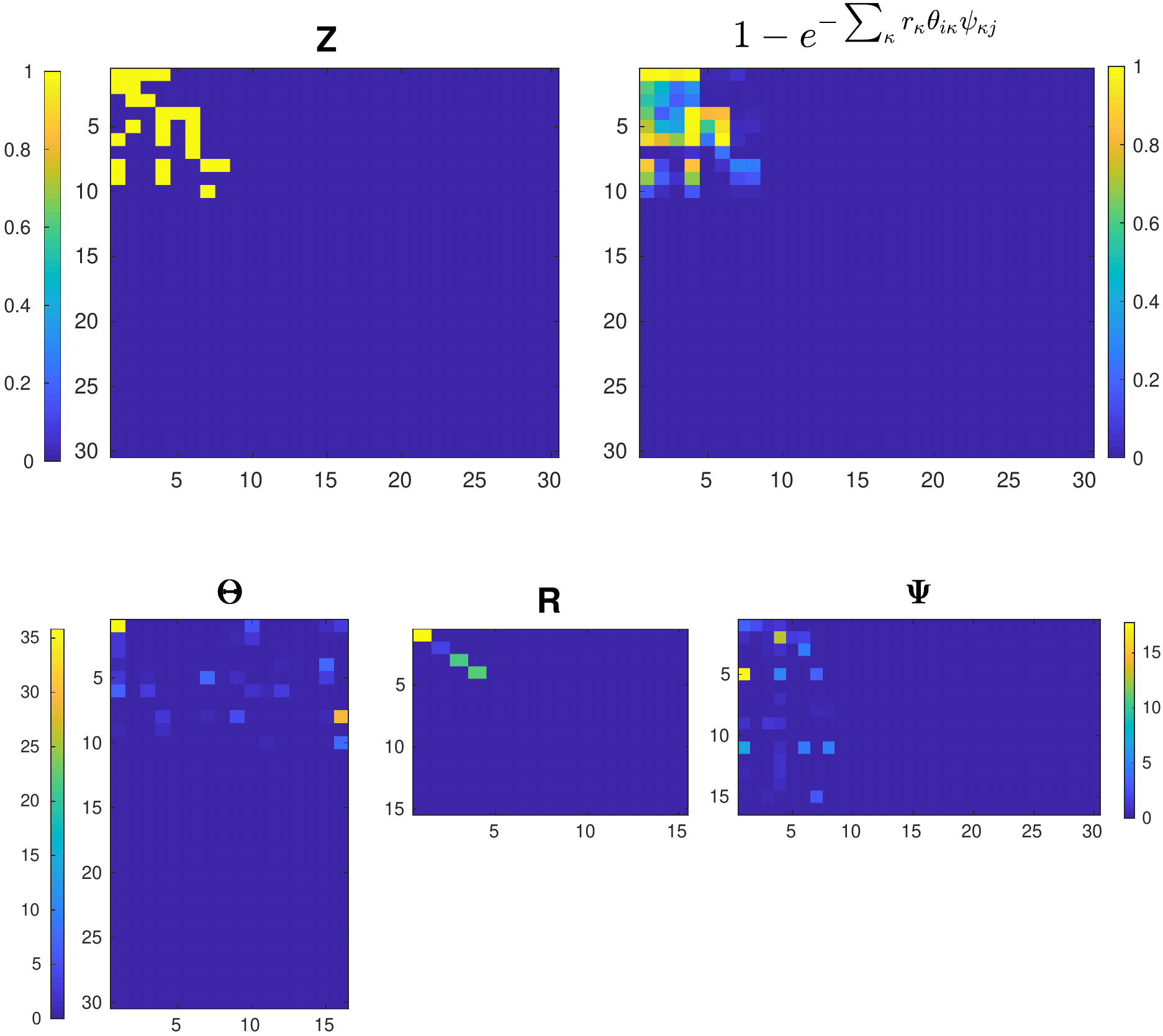}}\hspace{0em}%
\caption{\small }
  \label{fig:ZPThetaPsi_pedestrian}
       \end{center}
  \end{subfigure}
    \begin{subfigure}[b]{.5\linewidth}
     \begin{center}
\includegraphics[width=80mm]{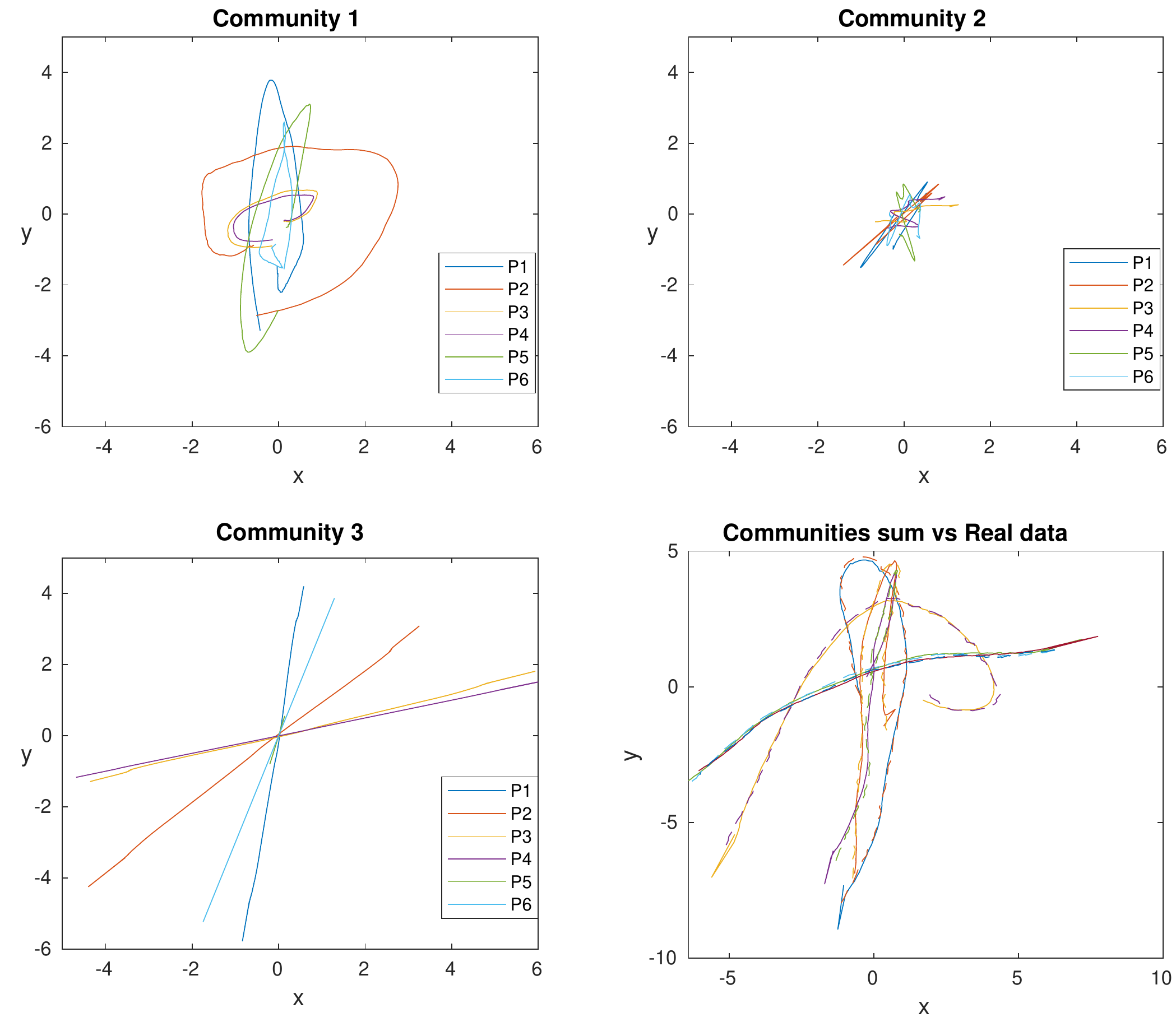}
   \caption{\small }
     \label{fig:pedestrianb}
          \end{center}
   \end{subfigure}\vspace{-3mm}
   \caption{\small Visualization of a posterior sample of  GGP-LDS applied to the Pedestrian dataset, where (a) and (b) are analogous plots to Figs. \ref{fig:ZPThetaPsi} and \ref{fig:lorenz2}, respectively.}
  \end{figure}
Fig.~\ref{fig:ZPThetaPsi_pedestrian} provides the interpretation  of the latent factors  for this dataset, analogous to Fig.~\ref{fig:ZPThetaPsi} used to provide interpretation of the latent structure inferred from  Lorenz Attractor. Fig. \ref{fig:pedestrianb}, analogous to Fig.~\ref{fig:lorenz2} for Lorenz Attractor, represents the reconstruction of all 6 pedestrians' trajectories using the three strongest communities. 
A total of  four communities is inferred by GGP-LDS to model the underlying pattern of the data. Fig.~\ref{fig:pedestrianb} shows how each community can decompose the observed data in 2 dimensions $(x,y)$ into a community-specific sub-sequence.
The last subplot superposes  the first three sub-sequences and compares it against the true trajectory (shown in dashed lines).

It is interesting to see how each community can create one type of motion ($e.g.$, straight movement, circular trajectory, and spiral movement). It is evident that regardless of the property of motion, such as ``turn direction,'' ``radius of circular motion,'' or ``direction of straight,'' the trajectories of the same nature have appeared in a same community-specific sub-sequence. It can be seen in the figure that some of the communities have a very small contribution for some of the pedestrians' trajectory reconstruction since those pedestrians did not use that specific pattern in their recorded walking.

\subsection{Stock Price}

\begin{table*}[t]
 \caption{\small Analogous table to Table \ref{Lorenz_prediction} for the Stock Price dataset}
  \label{stock12}
  \centering
  \resizebox{1\columnwidth}{!}{
  \begin{tabular}{lllllllllll}
    \toprule
    \multicolumn{11}{c}{Mean absolute error for 10 forecast horizons}                   \\
    \cmidrule(r){1-11}
    
    Algorithm & $t=1$ & $t=2$  & $t=3$  & $t=4$  & $t=5$   & $t=6$ & $t=7$  & $t=8$  & $t=9$  & $t=10$\\
    \midrule
    LDS 	   & $0.07_{(0.01)}$ 	&$0.06_{(0.02)}$	 	&$0.07_{(0.01)}$		&$0.07_{(0.02)}$	&$0.08_{(0.02)}$   & $0.08_{(0.02)}$ & $0.09_{(0.02)}$ & $0.08_{(0.02)}$ & $0.09_{(0.03)}$ & $0.10_{(0.02)}$     \\
    
    rLDSg     		   & $0.03_{(0.07)}$ 	&$0.03_{(0.01)}$	 	&$0.04_{(0.19)}$	&$0.04_{(0.01)}$	&$0.06_{(0.01)}$    & $0.06_{(0.02)}$ & $0.06_{(0.02)}$ & $0.07_{(0.02)}$ & $0.08_{(0.02)}$ & $0.09_{(0.02)}$   \\

    SGLDS    		  & $0.05_{(0.01)}$ & $0.05_{(0.01)}$ & $0.06_{(0.02)}$ & $0.06_{(0.02)}$ & $0.06_{(0.03)}$    & $0.08_{(0.02)}$ & $0.08_{(0.03)}$ & $0.09_{(0.03)}$ & $0.09_{(0.03)}$ & $0.10_{(0.04)}$  \\
    
    TrSLDS     		& $0.03_{(0.01)}$ & $0.04_{(0.01)}$ & $0.04_{(0.02)}$ & $\mathbf{0.05_{(0.02)}}$ & $0.06_{(0.03)}$    & $0.07_{(0.02)}$ & $0.07_{(0.03)}$ & $0.08_{(0.03)}$ & $0.08_{(0.03)}$ & $0.09_{(0.03)}$  \\

    Multi-output GP 	  & $0.09_{(0.03)}$ 	&$0.08_{(0.03)}$	 	&$0.10_{(0.03)}$	&$0.13_{(0.03)}$	&$0.11_{(0.03)}$  & $0.10_{(0.02)}$ & $0.12_{(0.03)}$ & $0.12_{(0.03)}$ & $0.13_{(0.03)}$ & $0.14_{(0.03)}$ \\
    
    FB Prophet     & $\mathbf{0.01_{(0.02)}}$	 &$\mathbf{0.03_{(0.01)}}$	 	&$0.06_{(0.01)}$	&$0.08_{(0.02)}$	&$0.07_{(0.01)}$  & $0.08_{(0.02)}$ & $0.09_{(0.02)}$ & $0.08_{(0.02)}$ & $0.09_{(0.03)}$ & $0.10_{(0.02)}$  \\
    
    DeepAR 	  & $0.05_{(0.01)}$ & $0.05_{(0.02)}$ & $0.06_{(0.02)}$ & $0.06_{(0.03)}$ & $0.07_{(0.02)}$  & $0.08_{(0.03)}$ & $0.08_{(0.03)}$ & $0.09_{(0.03)}$ & $0.09_{(0.04)}$ & $0.10_{(0.03)}$\\
    
    TRCRP     & $0.03_{(0.01)}$ 	&$0.04_{(0.02)}$	 	&$0.04_{(0.01)}$		&${0.06_{(0.01)}}$	&${0.07_{(0.02)}}$  &$\mathbf{0.05_{(0.02)}}$	&$\mathbf{0.04_{(0.02)}}$	&$\mathbf{0.03_{(0.01)}}$	&$\mathbf{0.05_{(0.01)}}$	&$0.07_{(0.02)}$ \\
    
      GGP-LDS (10 steps)     &$0.03_{(0.01)}$	&$\mathbf{0.03_{(0.01)}}$	&$\mathbf{0.03_{(0.01)}}$	&$0.05_{(0.02)}$	&$\mathbf{0.05_{(0.01)}}$ &$\mathbf{0.05_{(0.01)}}$	&$0.05_{(0.01)}$	&$0.05_{(0.02)}$	&$0.06_{(0.02)}$	&$\mathbf{0.07_{(0.02)}}$\\

  \cmidrule(r){2-11}  
  
   GGP-LDS (1 step)	  &$0.02_{(0.01)}$	&$0.02_{(0.01)}$	&$0.02_{(0.01)}$	&$0.03_{(0.01)}$	&$0.02_{(0.01)}$ &$0.02_{(0.01)}$	&$0.02_{(0.01)}$	&$0.02_{(0.01)}$	&$0.04_{(0.01)}$	&$0.05_{(0.02)}$\\

    \bottomrule
  \end{tabular}}%
\end{table*}

\begin{figure}[t] 
  \centering
  \subcaptionbox{\label{figstocka}}{\includegraphics[width=75mm]{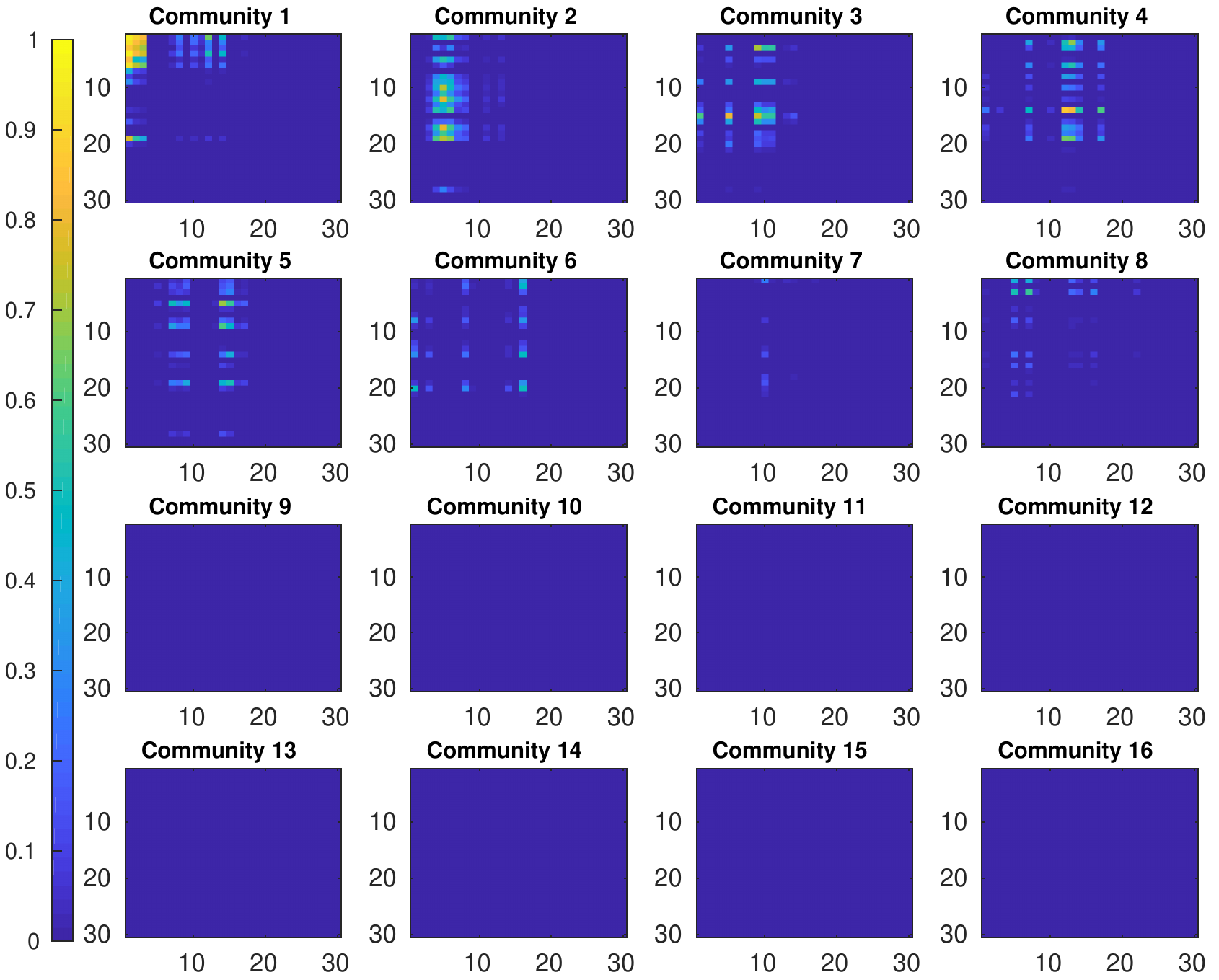}} %
\vspace{-0mm}
  \subcaptionbox{\label{figstockb}}{\includegraphics[width=85mm]{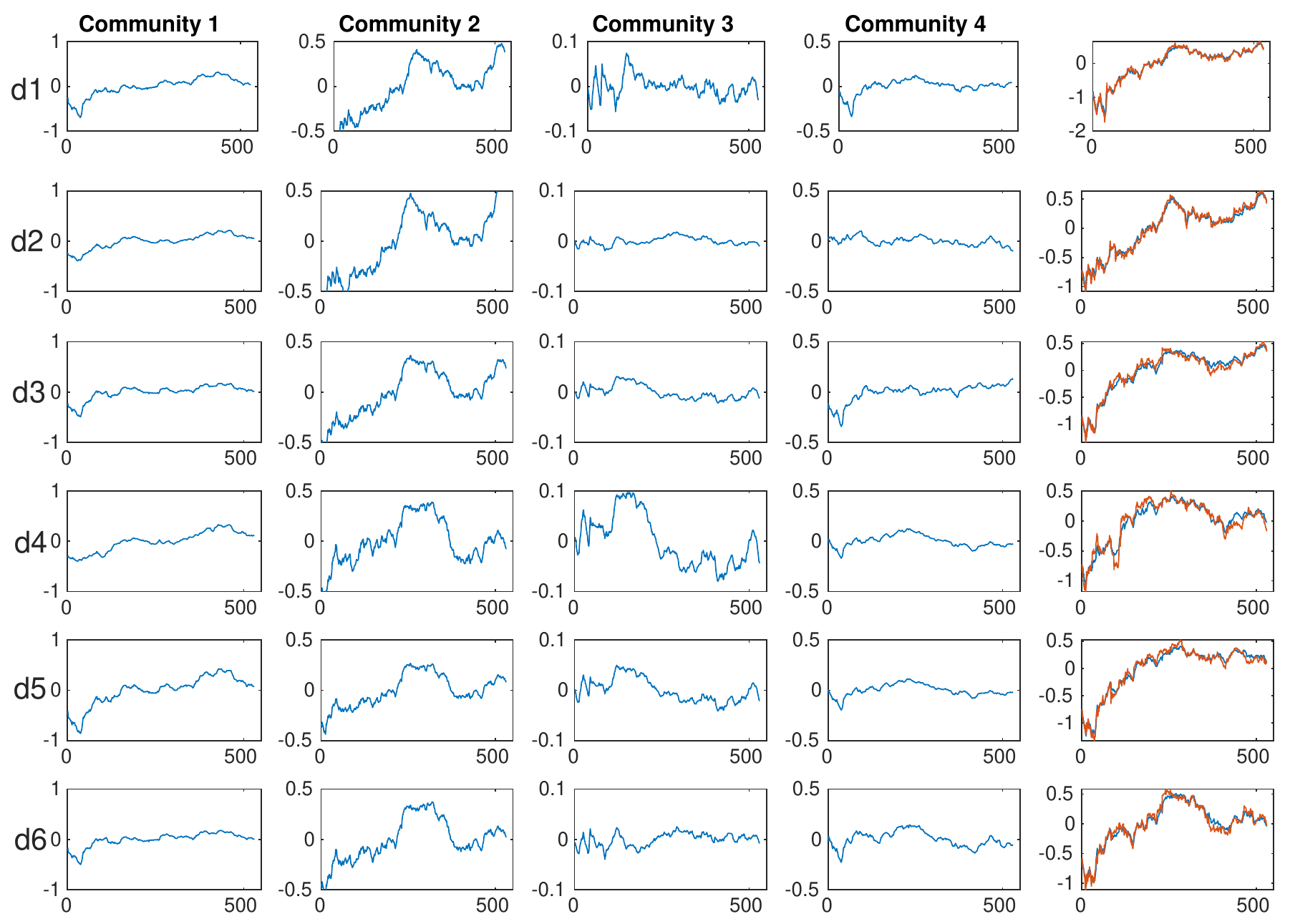}}\vspace{-0mm}
\caption{\small Visualization of GGP-LDS applied to stock closing  prices. (a) Inferred communities, analogous plot to Fig. \ref{fig1_b}  (b) Inferred community-specific sub-sequences of the four strongest communities and their superpositions, analogous plot to Fig. \ref{fig:lorenz2}.}
  \label{fig:Stock}
\end{figure}

This dataset contains 12 companies' relative closing price ($P_t-P_{t-1}$) over the course of three years. These 12 companies have been selected from four different industries, and the stock closing prices of the ones in the same industry share similar temporal behaviors. %
Table \ref{stock12} compares the predictive performance of various algorithms  on this dataset. In most of the 10 forecast horizons our model has outperformed the other competing models. The parameter settings of various algorithms are the same as those in Section \ref{sec:Lorenz} for Lorenz Attractor, with the following exceptions:
\begin{itemize}
\item LDS of \citet{ghahramani96parameterestimation}, inferred by EM, with $K=12$.
\item $rLDS_g$  of \citet{liu2015regularized}, with $K=14$ and random initialization.
\item TrSLDS of \citet{nassar2018tree}, with its tree depth set as 1 and dimension of latent states for each LDS as 8.
\end{itemize}

Fig. \ref{figstocka} shows the formed communities. Eight non-zero communities has been formed with the first four communities having at least one non-overlapping member, while the next four communities do not have members that are exclusive to them. %
Having four major communities, Fig. \ref{figstockb} shows how each of these major communities can contribute to reconstruct the observed data. Rows of Fig. \ref{figstockb} correspond to 6 stocks out of 12. The first four columns of the figure describe how the four strongest detected communities contributed to reconstruct the data in each dimension. There are two noticeable observations in Fig. \ref{figstockb}. First, it can be seen that each community represents similar behavior for all 6 selected stocks in the figure, and these behaviors are distinct from one community to another. Second, it is evident that if a behavior represented by a community does not play a significant role in reconstruction of the data in a specific dimension, that community contribution will be trivial. As an example, community 3's role in reconstructing the second stock is trivial, while playing a much more significant role to reconstruct the observation of the fourth stock. 

\subsection{Daily new COVID-19 Cases}

This dataset contains the daily reported COVID-19 cases in each of the 50 States and District of Columbia from the beginning of March to end of May 2020. For predictive performance comparison, we have compared our model against two models which have been specifically developed to understand the COVID-19 pandemic: \citet{zou2020epidemic} use a differential equation based epidemic model and use differential equation solver to find the model parameters; \citet{woody2020projections} utilize the social distancing data as covariates and a statistical model negative binomial regression model utilizing these covariates to predict the intensity of pandemic %
in the future. 
Table \ref{tbl:covid19} shows the $6$-steps predictive performance of GGP-GLDS on this dataset. For the competing algorithms in Table \ref{tbl:covid19}, we list the parameter settings below:
\begin{itemize}
\item TrSLDS of \citet{nassar2018tree}, with its tree depth set as 1 and dimension of latent states for each LDS as 5.
\item TRCRP of \citet{saad2018temporally}, with the Markov chain order set as $p=6$ .
\item UCLA-SuEIR of \citet{zou2020epidemic} 
\item UT-Mobility of \citet{woody2020projections} 
\end{itemize}

\begin{table*}[t!]
   \caption{\small Prediction of daily  new Covid-19 cases in the U.S. given daily observations from March 6 to May 23, 2020. }
  \label{stock12}
  \centering
  \resizebox{.8\columnwidth}{!}{
  \begin{tabular}{lllllll}
    \toprule
    \multicolumn{7}{c}{Mean absolute relative error for 6 forecast horizons}                   \\
    \cmidrule(r){1-7}
    
    Algorithm & $t=1$ & $t=2$  & $t=3$  & $t=4$  & $t=5$   & $t=6$ \\
   Year 2020 & May 24 & May 25  & May 26  & May 27  & May 28   & May 29\\
    \midrule

    TrSLDS     		& $6.68_{(2.31)}$ & $7.21_{(3.28)}$ & $6.18_{(3.08)}$ & ${5.25_{(2.42)}}$ & $1.87_{(2.8)}$    & $8.85_{(3.27)}$ \\

    UCLA Covid19     & $\mathbf{3.87_{(1.68)}}$	 &${4.86_{(1.51)}}$	 	&$5.23_{(2.25)}$	&$3.23_{(2.27)}$	&$2.87_{(1.96)}$  & $6.61_{(4.18)}$  \\
    
    UT-Austin Covid19 	  & $6.32_{(2.51)}$ & $5.38_{(2.73)}$ & $6.26_{(3.25)}$ & $3.78_{(2.13)}$ & $3.16_{(2.02)}$  & $7.95_{(3.93)}$ \\
    
    TRCRP     & $7.66_{(3.81)}$ 	&$6.27_{(3.25)}$	 	&$7.22_{(3.36)}$		&${4.21_{(2.86)}}$	&${2.31_{(1.37)}}$  &$\mathbf{6.11_{(3.25)}}$ \\
    
      GGP-GLDS (6 steps)     &$4.78_{(2.95)}$	&$\mathbf{4.47_{(2.27)}}$	&$\mathbf{3.71_{(1.12)}}$	&$\mathbf{1.68_{(5.67)}}$	&$\mathbf{0.69_{(0.42)}}$ &${9.50_{(18.01)}}$\\

    \bottomrule
  \end{tabular}}\label{tbl:covid19}
\end{table*}

 \begin{figure}[t]
    \centering
    \includegraphics[width=115mm]{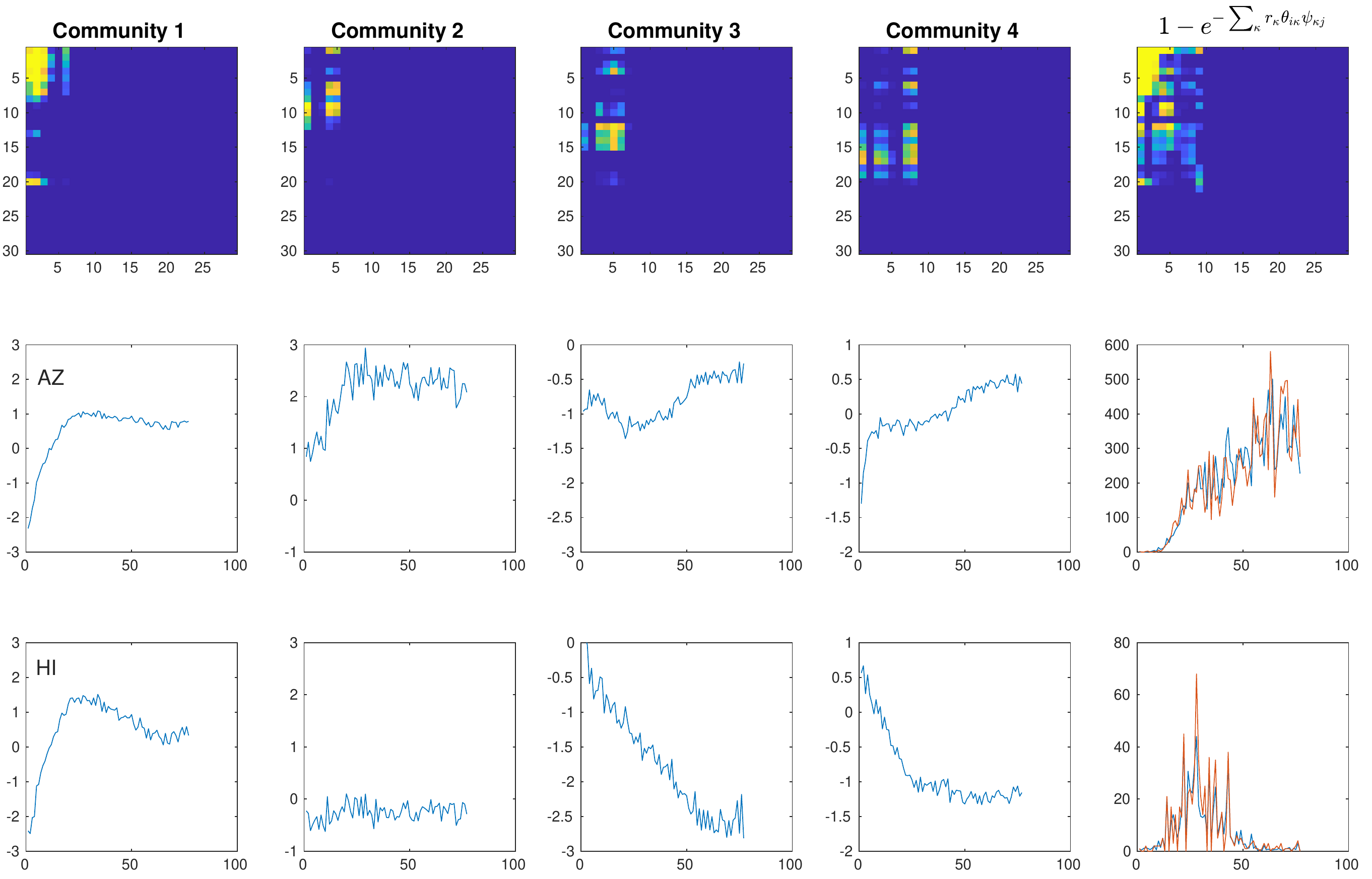}
  \caption{\small Visualization of GGP-GLDS (NBDS) applied to U.S. COVID-19 daily cases. %
  Row 1 shows the relative activation strengths of the four strongest communities, analogous to Fig. \ref{fig1_b};
  Rows 2 and 3 show the inferred community-specific sub-sequences of the four strongest communities and the expected counts given their superpositions.} %
 \vspace{-2mm}
 \label{fig:Covid}
\end{figure}

Similar to previous visualizations,  %
the first row of Fig.~\ref{fig:Covid} shows the relative strengths of four discovered communities that have at least one non-overlapping member, with the last column of the first row representing $1-e^{-\sum_{\kappa}r_\kappa \theta_{i\kappa} \psi_{j\kappa}}$. The relative strength of a community is defined as in \eqref{eq:relativestrength}. The second row shows the trajectories for Arizona (AZ), with the trajectory in each of its first four columns showing a community-specific sub-sequence for AZ, which is the corresponding row of  
 $%
  \Dmat\hat{\xv}_{1:T}^{(\kappa)}$, where $\hat{\xv}_{t}^{(\kappa)} = [(\Wmat\odot \Zmat)\odot \Amat_\kappa]\xv_{t-1}$ is defined as in \eqref{eq:subseq};
the last column of the second row compares 
 the trajectory obtained by $$%
 \hat{\yv}_{t}^{(\kappa)} :=\eta e^{(\sum_{\kappa=1}^4 \Dmat\hat{\xv}_{t}^{(\kappa)})}$$ against the true trajectory of AZ, %
 which shows how discovered communities can reconstruct the observed time series at different data dimensions. 
Shown in the third row are analogous plots for Hawaii (HI).

\begin{figure}[t] 
\begin{center}
 \begin{subfigure}[b]{.48\linewidth}
  \hspace{-3mm}\includegraphics[width=82mm]{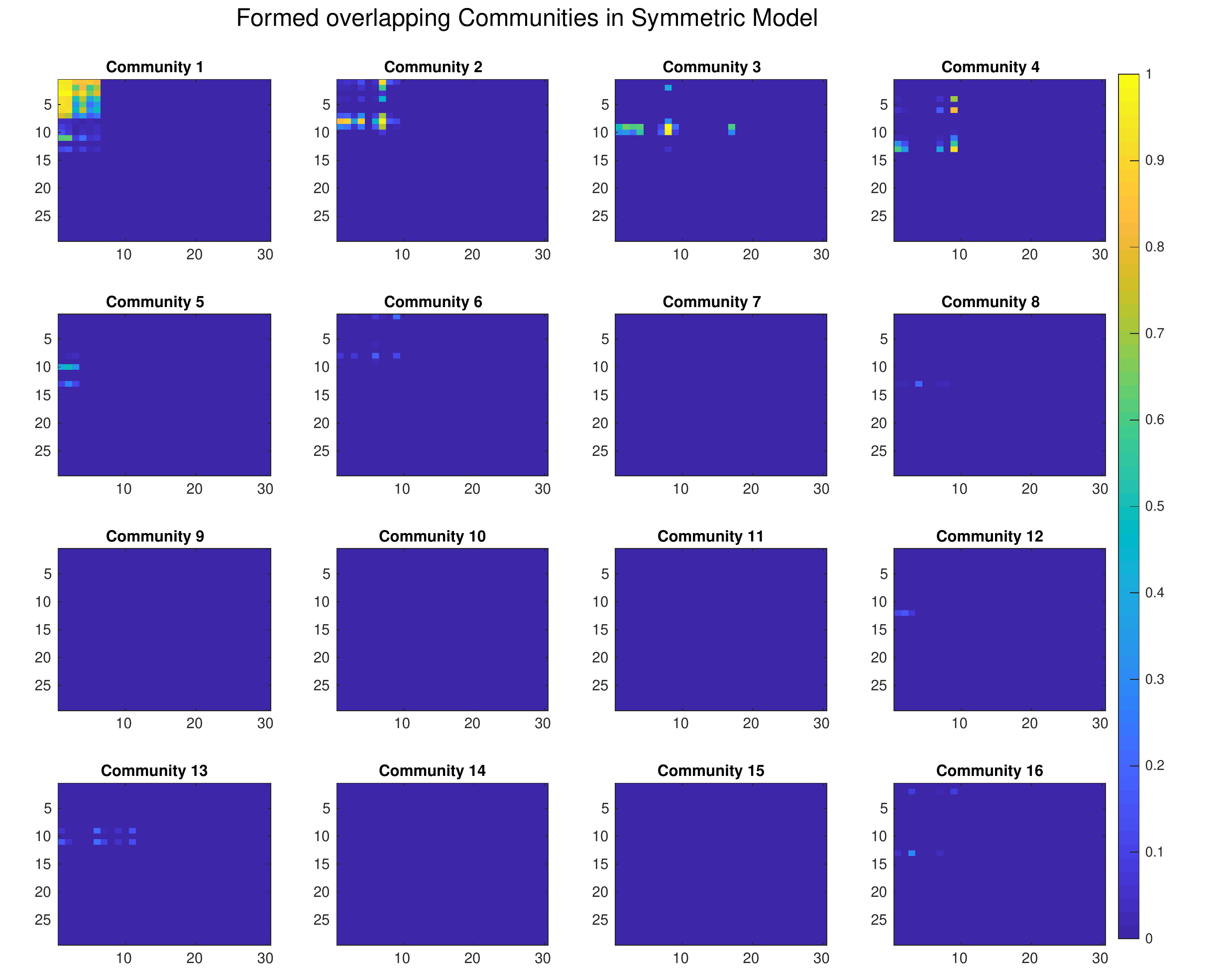}
  \caption{\small }
   \label{fig:covid_communities}
  \end{subfigure}\quad
   \begin{subfigure}[b]{.48\linewidth}
  \includegraphics[width=83mm,height=65mm]{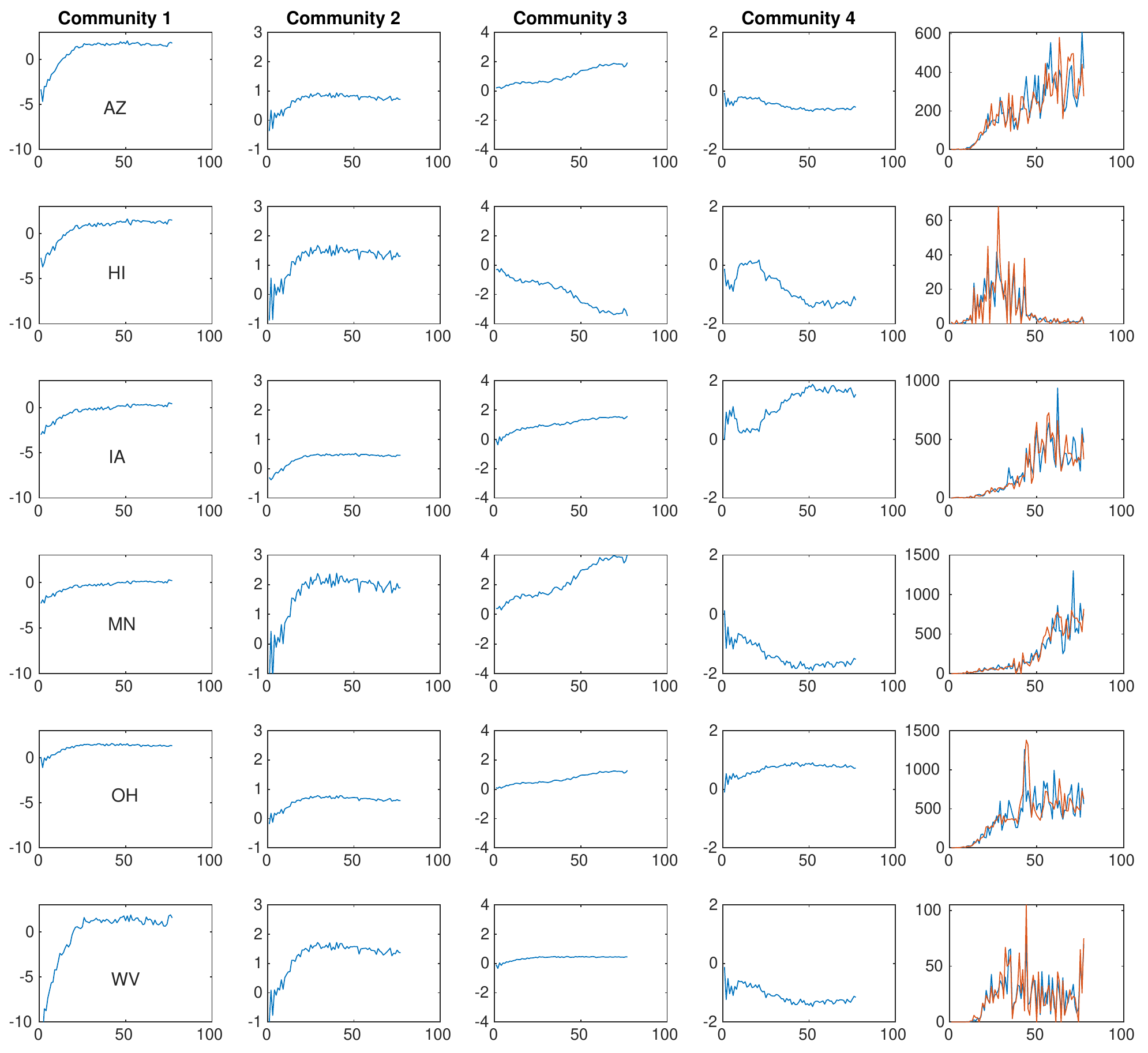}
\caption{\small }
\label{fig:covid2}
  \end{subfigure}
\caption{\small 
Visualization of a posterior sample (different from the sample used for  Fig. \ref{fig:Covid}) of GGP-GLDS (NBDS) applied to U.S. daily new COVID-19 cases. (a): Inferred communities, analogous to Fig. \ref{fig1_b};  (b): Inferred community-specific sub-sequences of the four strongest communities and their superpositions, analogous to Fig. \ref{fig:lorenz2}.
}
 \label{fig:Covid_ZC}
 \end{center}
\end{figure}

Fig.~\ref{fig:covid_communities} shows all discovered communities,  out of  which the first four communities have at least one non-overlapping member unique to themselves.
 Please note that the discovered communities and adjacency matrix of the posterior sample in Fig. \ref{fig:Covid_ZC}
 are different from those in Fig.~\ref{fig:Covid}. We intentionally choose two different samples from 2 different runs with different random seeds to show the consistency of interpretation across different runs. Although these two posterior samples come from  two independent Markov chains initialized with different random seeds, and consequently different adjacency matrices, the algorithm still finds 4 communities with at least one non-overlapping member, and Figs. \ref{fig:covid2} and \ref{fig:Covid} show that the interpretation of the communities stay consistent from one run to another. This consistency of interpretation helps to show that no specific method for model parameter initialization is required for the proposed GGP-GLDS.  %
 
 Similar to previous visualizations, there are two noticeable observations from Figs.~\ref{fig:Covid} and \ref{fig:covid2}. First, it can be seen how  each community represents a distinct behavior shared by all data dimensions at different magnitudes. Second, a behavior represented by a community may play a negligible role in reconstructing the data in a specific dimension if its corresponding magnitude is small.

\section{Discussion and Conclusion}
We introduce the graph gamma process (GGP) to form an infinite dimensional transition model with a finite random number of nonzero-degree nodes and a finite random number of nonzero-edge communities over these nodes. The GGP is used to promote sparsity on the state-transition matrix of a (generalized) linear dynamical system, and encourage forming overlapping communities among the nonzero-degree nodes of the graph. 
The model is designed %
such that each node community models a behavior described with a linear dynamical system. Instead of assigning one behavior to a temporal segment of an observation trajectory, it allows 
the whole time series 
to be viewed as a superposition of distinctly behaved trajectories, each of which is modeled by one of the discovered overlapping communities. These overlapping communities can be activated at different levels at different time points to support the superposition to have sophisticated temporal behavior, allowing smooth transitions between different linear dynamics to approximate nonlinear dynamical behaviors.  
This paper focuses on time-series data modeling using a  nonparametric Bayesian hierarchical model. The main challenges for modeling the time series include accurate forecast, missing data imputation, and meaningful clustering of the latent variables of the model such that these clusters can be translated to meaningful underlying patterns of the data, which represent different behaviors in a dataset without requiring data specific customization. The main goal of this paper is to develop a nonparametric Bayesian model to find a flexible solution that provides interpretable latent representation and have good predictive performance. The nonparametric Bayesian construction of the model introduced in this paper  
prevents the underlying model from being
over- and under-parameterized, which can result in over- or under-fitting, respectively, and hence poor performance in forecasting. In addition, the model learns a parsimonious graph among the infinite dimensional latent variables such that the induced clusters by community forming among non-zero-degree states can explain the underlying patterns of the data without the need of manual tuning. The nonparametric Bayesian construction and shrinkage property of the GGP introduced in this paper avoid inducing too many latent communities over too many latent states %
that will make the model difficult to interpret. Our experiments show that GGP-(G)LDS creates good balance between interpretability and predictive performance. Using the formed overlapping communities over the inferred set of latent states, the model is capable of modeling non-linearity in the data by creating multiple linear systems which try to learn the underlying non-linear behavior of the data. The model can break the sophisticated behavior in a trajectory to the combination of simpler behaviors modeled by linear dynamical systems, and it helps to model the nonlinearity of the data using multiple linear dynamical systems and smooth transition between them. The linear approximation of data with nonlinear dynamics enables the proposed GGP-LDS to leverage existing smoothing and filtering algorithms readily available to canonical LDSs.

Furthermore, this model can be seen as a generalization to existing nonparametric Bayesian linear dynamical systems from two different perspectives. %
First, the formed communities with overlapping members will translate to the soft switching concept among latent communities (or different LDSs) such that the strength of each LDS at any given time can change using the overlapping states which have replaced the hard switching among the different LDSs in switching LDSs. The hard switch is prone to picking a false LDS  at the switching point, and results in wrong interpretation and deteriorated predictive performance. Second, unlike switching LDSs, the proposed model does not make any assumption on the dimensions of latent states, which are often set to be different in switching LDSs for different segments of the trajectory.

In addition, we show by an example how the proposed model with minor change in observation layer can turn to a model for generalized linear models (GLM) which helps us to model overdispersed count data, which often seriously violates the Gaussian assumption on observations. That adaptability helps us to model the trajectories of daily new  COVID-19 cases in each State of the US by training the model on  multivariate count observations across the 50 States and D.C., and hope its interpretability and predictive performance could help decision makers in  pattern discovery, resource assignment, and predictive analytics related tasks. An on-going effort is using the proposed negative binomial dynamical system to jointly model the number of daily COVID-19 cases and deaths across all the States at the U.S. and provide forecast for future COVID-19 cases and deaths.

\bibliography{GGP}

\begin{thebibliography}{51}
\providecommand{\natexlab}[1]{#1}
\providecommand{\url}[1]{\texttt{#1}}
\expandafter\ifx\csname urlstyle\endcsname\relax
  \providecommand{\doi}[1]{doi: #1}\else
  \providecommand{\doi}{doi: \begingroup \urlstyle{rm}\Url}\fi

\bibitem[Alvarez and Lawrence(2009)]{alvarez2009sparse}
M.~Alvarez and N.~D. Lawrence.
\newblock Sparse convolved gaussian processes for multi-output regression.
\newblock In \emph{Advances in neural information processing systems}, pages
  57--64, 2009.

\bibitem[Barber et~al.(2011)Barber, Cemgil, and Chiappa]{barber2011bayesian}
D.~Barber, A.~T. Cemgil, and S.~Chiappa.
\newblock \emph{Bayesian time series models}.
\newblock Cambridge University Press, 2011.

\bibitem[Caron et~al.(2008)Caron, Davy, Doucet, Duflos, and
  Vanheeghe]{caron2008noise}
F.~Caron, M.~Davy, A.~Doucet, E.~Duflos, and P.~Vanheeghe.
\newblock Bayesian inference for linear dynamic models with dirichlet process
  mixtures.
\newblock \emph{IEEE Transactions on Signal Processing}, 56\penalty0
  (1):\penalty0 71--84, 2008.

\bibitem[Carvalho and Lopes(2007)]{carvalho2007simulation}
C.~M. Carvalho and H.~F. Lopes.
\newblock Simulation-based sequential analysis of {Markov} switching stochastic
  volatility models.
\newblock \emph{Computational Statistics \& Data Analysis}, 51\penalty0
  (9):\penalty0 4526--4542, 2007.

\bibitem[Charles et~al.(2011)Charles, Asif, Romberg, and
  Rozell]{charles2011sparsity}
A.~Charles, M.~S. Asif, J.~Romberg, and C.~Rozell.
\newblock Sparsity penalties in dynamical system estimation.
\newblock In \emph{Information Sciences and Systems (CISS), 2011 45th Annual
  Conference on}, pages 1--6. IEEE, 2011.

\bibitem[Chiuso and Pillonetto(2010)]{chiuso2010learning}
A.~Chiuso and G.~Pillonetto.
\newblock Learning sparse dynamic linear systems using stable spline kernels
  and exponential hyperpriors.
\newblock In \emph{Advances in Neural Information Processing Systems}, pages
  397--405, 2010.

\bibitem[Davis et~al.(2016)Davis, Zang, and Zheng]{davis2016sparse}
R.~A. Davis, P.~Zang, and T.~Zheng.
\newblock Sparse vector autoregressive modeling.
\newblock \emph{Journal of Computational and Graphical Statistics}, 25\penalty0
  (4):\penalty0 1077--1096, 2016.

\bibitem[Flunkert et~al.(2017)Flunkert, Salinas, and
  Gasthaus]{flunkert2017deepar}
V.~Flunkert, D.~Salinas, and J.~Gasthaus.
\newblock Deepar: Probabilistic forecasting with autoregressive recurrent
  networks.
\newblock \emph{arXiv preprint arXiv:1704.04110}, 2017.

\bibitem[Fox et~al.(2009)Fox, Sudderth, Jordan, and Willsky]{FoxSLDS09}
E.~Fox, E.~B. Sudderth, M.~I. Jordan, and A.~S. Willsky.
\newblock Nonparametric bayesian learning of switching linear dynamical
  systems.
\newblock In \emph{NIPS}, pages 457--464. 2009.

\bibitem[Gao et~al.(2016)Gao, Archer, Paninski, and Cunningham]{gao2016linear}
Y.~Gao, E.~W. Archer, L.~Paninski, and J.~P. Cunningham.
\newblock Linear dynamical neural population models through nonlinear
  embeddings.
\newblock In \emph{Advances in neural information processing systems}, pages
  163--171, 2016.

\bibitem[Ghahramani and Hinton(1996)]{ghahramani96parameterestimation}
Z.~Ghahramani and G.~E. Hinton.
\newblock Parameter estimation for linear dynamical systems.
\newblock Technical report, 1996.

\bibitem[Ghahramani and Roweis(1999)]{ghahramani1999learning}
Z.~Ghahramani and S.~T. Roweis.
\newblock Learning nonlinear dynamical systems using an {EM} algorithm.
\newblock In \emph{NIPS}, pages 431--437, 1999.

\bibitem[Hardt et~al.(2018)Hardt, Ma, and Recht]{hardt2018gradient}
M.~Hardt, T.~Ma, and B.~Recht.
\newblock Gradient descent learns linear dynamical systems.
\newblock \emph{The Journal of Machine Learning Research}, 19\penalty0
  (1):\penalty0 1025--1068, 2018.

\bibitem[Harrison et~al.(2003)Harrison, Penny, and
  Friston]{harrison2003multivariate}
L.~Harrison, W.~D. Penny, and K.~Friston.
\newblock Multivariate autoregressive modeling of fmri time series.
\newblock \emph{Neuroimage}, 19\penalty0 (4):\penalty0 1477--1491, 2003.

\bibitem[Hernandez et~al.(2018)Hernandez, Moretti, Wei, Saxena, Cunningham, and
  Paninski]{hernandez2018novel}
D.~Hernandez, A.~K. Moretti, Z.~Wei, S.~Saxena, J.~Cunningham, and L.~Paninski.
\newblock A novel variational family for hidden nonlinear markov models.
\newblock \emph{arXiv preprint arXiv:1811.02459}, 2018.

\bibitem[Imani and Braga-Neto(2018)]{imani2018particle}
M.~Imani and U.~M. Braga-Neto.
\newblock Particle filters for partially-observed boolean dynamical systems.
\newblock \emph{Automatica}, 87:\penalty0 238--250, 2018.

\bibitem[Ishwaran and Rao(2005)]{ishwaran2005spike}
H.~Ishwaran and J.~S. Rao.
\newblock Spike and slab variable selection: frequentist and {B}ayesian
  strategies.
\newblock \emph{Annals of statistics}, pages 730--773, 2005.

\bibitem[Johnson et~al.(2016)Johnson, Duvenaud, Wiltschko, Adams, and
  Datta]{johnson2016composing}
M.~Johnson, D.~K. Duvenaud, A.~Wiltschko, R.~P. Adams, and S.~R. Datta.
\newblock Composing graphical models with neural networks for structured
  representations and fast inference.
\newblock In \emph{Advances in neural information processing systems}, pages
  2946--2954, 2016.

\bibitem[Johnson and Willsky(2013)]{johnson2013bayesian}
M.~J. Johnson and A.~S. Willsky.
\newblock Bayesian nonparametric hidden semi-markov models.
\newblock \emph{Journal of Machine Learning Research}, 14\penalty0
  (Feb):\penalty0 673--701, 2013.

\bibitem[Kalantari et~al.(2018)Kalantari, Ghosh, and
  Zhou]{Kalantari2018nonparametric}
R.~Kalantari, J.~Ghosh, and M.~Zhou.
\newblock Nonparametric bayesian sparse graph linear dynamical systems.
\newblock \emph{arXiv preprint arXiv:1802.07434}, 2018.

\bibitem[Kalman(1960)]{Kalman}
R.~E. Kalman.
\newblock A new approach to linear filtering and prediction problems.
\newblock \emph{Transactions of the ASME--Journal of Basic Engineering},
  82\penalty0 (Series D):\penalty0 35--45, 1960.

\bibitem[Kalman(1963)]{kalman1963mathematical}
R.~E. Kalman.
\newblock Mathematical description of linear dynamical systems.
\newblock \emph{Journal of the Society for Industrial and Applied Mathematics,
  Series A: Control}, 1\penalty0 (2):\penalty0 152--192, 1963.

\bibitem[Kennedy and Turley(2011)]{kennedy2011time}
C.~E. Kennedy and J.~P. Turley.
\newblock Time series analysis as input for clinical predictive modeling:
  Modeling cardiac arrest in a pediatric icu.
\newblock \emph{Theoretical Biology and Medical Modelling}, 8\penalty0
  (1):\penalty0 40, 2011.

\bibitem[Kingman(1993)]{poissonp}
J.~F.~C. Kingman.
\newblock \emph{Poisson Processes}.
\newblock Oxford University Press, 1993.

\bibitem[Koyama(2018)]{koyama2018projection}
S.~Koyama.
\newblock Projection smoothing for continuous and continuous-discrete
  stochastic dynamic systems.
\newblock \emph{Signal Processing}, 144:\penalty0 333--340, 2018.

\bibitem[Lai et~al.(2018)Lai, Chang, Yang, and Liu]{lai2018modeling}
G.~Lai, W.-C. Chang, Y.~Yang, and H.~Liu.
\newblock Modeling long-and short-term temporal patterns with deep neural
  networks.
\newblock In \emph{The 41st International ACM SIGIR Conference on Research \&
  Development in Information Retrieval}, pages 95--104. ACM, 2018.

\bibitem[Li et~al.(2010)Li, Hu, Xu, Xiao, and Wang]{li2010application}
J.~Li, C.~Hu, D.~Xu, J.~Xiao, and H.~Wang.
\newblock Application of time-series autoregressive integrated moving average
  model in predicting the epidemic situation of newcastle disease.
\newblock In \emph{World Automation Congress (WAC), 2010}, pages 141--144.
  IEEE, 2010.

\bibitem[Linderman et~al.(2017)Linderman, Johnson, Miller, Adams, Blei, and
  Paninski]{pmlr-v54-linderman17a}
S.~Linderman, M.~Johnson, A.~Miller, R.~Adams, D.~Blei, and L.~Paninski.
\newblock Bayesian learning and inference in recurrent switching linear
  dynamical systems.
\newblock In \emph{AISTATS}, volume~54, pages 914--922, Fort Lauderdale, FL,
  USA, 20--22 Apr 2017.

\bibitem[Liu and Hauskrecht(2015)]{liu2015regularized}
Z.~Liu and M.~Hauskrecht.
\newblock A regularized linear dynamical system framework for multivariate time
  series analysis.
\newblock In \emph{AAAI}, pages 1798--1804, 2015.

\bibitem[Ljung(1999)]{Ljung}
L.~Ljung.
\newblock \emph{System Identification: Theory for the User, 2nd edition}.
\newblock Prentice Hall, 1999.

\bibitem[Mitchell and Beauchamp(1988)]{mitchell1988bayesian}
T.~J. Mitchell and J.~J. Beauchamp.
\newblock Bayesian variable selection in linear regression.
\newblock \emph{Journal of the American Statistical Association}, 83\penalty0
  (404):\penalty0 1023--1032, 1988.

\bibitem[Murphy(1998)]{murphy1998switching}
K.~P. Murphy.
\newblock Switching {K}alman filters.
\newblock 1998.

\bibitem[Nassar et~al.(2018)Nassar, Linderman, Bugallo, and
  Park]{nassar2018tree}
J.~Nassar, S.~Linderman, M.~Bugallo, and I.~M. Park.
\newblock Tree-structured recurrent switching linear dynamical systems for
  multi-scale modeling.
\newblock \emph{arXiv preprint arXiv:1811.12386}, 2018.

\bibitem[Nieto-Barajas et~al.(2014)Nieto-Barajas, Contreras-Crist{\'a}n,
  et~al.]{nieto2014bayesian}
L.~E. Nieto-Barajas, A.~Contreras-Crist{\'a}n, et~al.
\newblock A bayesian nonparametric approach for time series clustering.
\newblock \emph{Bayesian Analysis}, 9\penalty0 (1):\penalty0 147--170, 2014.

\bibitem[Polson and Scott(2011)]{polson2011default}
N.~G. Polson and J.~G. Scott.
\newblock Default {B}ayesian analysis for multi-way tables: A data-augmentation
  approach.
\newblock \emph{arXiv preprint arXiv:1109.4180}, 2011.

\bibitem[Polson et~al.(2013)Polson, Scott, and Windle]{polson2013bayesian}
N.~G. Polson, J.~G. Scott, and J.~Windle.
\newblock Bayesian inference for logistic models using p{\'o}lya--gamma latent
  variables.
\newblock \emph{Journal of the American statistical Association}, 108\penalty0
  (504):\penalty0 1339--1349, 2013.

\bibitem[Saad and Mansinghka(2018)]{saad2018temporally}
F.~Saad and V.~Mansinghka.
\newblock Temporally-reweighted chinese restaurant process mixtures for
  clustering, imputing, and forecasting multivariate time series.
\newblock In \emph{International Conference on Artificial Intelligence and
  Statistics}, pages 755--764, 2018.

\bibitem[Siddiqi et~al.(2010)Siddiqi, Boots, and Gordon]{siddiqi2010reduced}
S.~Siddiqi, B.~Boots, and G.~Gordon.
\newblock Reduced-rank hidden markov models.
\newblock In \emph{Proceedings of the Thirteenth International Conference on
  Artificial Intelligence and Statistics}, pages 741--748, 2010.

\bibitem[St{\"a}dler et~al.(2013)St{\"a}dler, Mukherjee,
  et~al.]{stadler2013penalized}
N.~St{\"a}dler, S.~Mukherjee, et~al.
\newblock Penalized estimation in high-dimensional hidden markov models with
  state-specific graphical models.
\newblock \emph{The Annals of Applied Statistics}, 7\penalty0 (4):\penalty0
  2157--2179, 2013.

\bibitem[Taylor and Letham(2018)]{taylor2018forecasting}
S.~J. Taylor and B.~Letham.
\newblock Forecasting at scale.
\newblock \emph{The American Statistician}, 72\penalty0 (1):\penalty0 37--45,
  2018.

\bibitem[Teh et~al.(2006)Teh, Jordan, Beal, and Blei]{HDP}
Y.~W. Teh, M.~I. Jordan, M.~J. Beal, and D.~M. Blei.
\newblock Hierarchical {D}irichlet processes.
\newblock \emph{J. Amer. Statist. Assoc.}, 101, 2006.

\bibitem[Tipping(2001)]{tipping2001sparse}
M.~E. Tipping.
\newblock Sparse bayesian learning and the relevance vector machine.
\newblock \emph{Journal of machine learning research}, 1\penalty0
  (Jun):\penalty0 211--244, 2001.

\bibitem[West and Harrison(1997)]{West:1997:BFD:261170}
M.~West and J.~Harrison.
\newblock \emph{Bayesian Forecasting and Dynamic Models (2Nd Ed.)}.
\newblock Springer-Verlag New York, Inc., New York, NY, USA, 1997.
\newblock ISBN 0-387-94725-6.

\bibitem[Woody et~al.(2020)Woody, Tec, Dahan, Gaither, Lachmann, Fox, Meyers,
  and Scott]{woody2020projections}
S.~Woody, M.~G. Tec, M.~Dahan, K.~Gaither, M.~Lachmann, S.~Fox, L.~A. Meyers,
  and J.~G. Scott.
\newblock Projections for first-wave covid-19 deaths across the us using
  social-distancing measures derived from mobile phones.
\newblock \emph{medRxiv}, 2020.

\bibitem[Zhang et~al.(2017)Zhang, Ayoub, and Sundaram]{zhang2017sensor}
H.~Zhang, R.~Ayoub, and S.~Sundaram.
\newblock Sensor selection for {K}alman filtering of linear dynamical systems:
  Complexity, limitations and greedy algorithms.
\newblock \emph{Automatica}, 78:\penalty0 202--210, 2017.

\bibitem[Zhou(2015)]{epm_aistats2015}
M.~Zhou.
\newblock Infinite edge partition models for overlapping community detection
  and link prediction.
\newblock In \emph{AISTATS}, pages 1135--1143, 2015.

\bibitem[Zhou(2016)]{zhou2016softplus}
M.~Zhou.
\newblock Softplus regressions and convex polytopes.
\newblock \emph{arXiv preprint arXiv:1608.06383}, 2016.

\bibitem[Zhou and Carin(2013)]{zhou2013negative}
M.~Zhou and L.~Carin.
\newblock Negative binomial process count and mixture modeling.
\newblock \emph{IEEE Transactions on Pattern Analysis and Machine
  Intelligence}, 37\penalty0 (2):\penalty0 307--320, 2013.

\bibitem[Zhou et~al.(2009)Zhou, Chen, Paisley, Ren, Sapiro, and
  Carin]{Mingyuan09}
M.~Zhou, H.~Chen, J.~Paisley, L.~Ren, G.~Sapiro, and L.~Carin.
\newblock Non-parametric {B}ayesian dictionary learning for sparse image
  representations.
\newblock In \emph{NIPS}, 2009.

\bibitem[Zhou et~al.(2012)Zhou, Li, Dunson, and Carin]{zhou2012lognormal}
M.~Zhou, L.~Li, D.~Dunson, and L.~Carin.
\newblock Lognormal and gamma mixed negative binomial regression.
\newblock In \emph{Proceedings of the... International Conference on Machine
  Learning. International Conference on Machine Learning}, volume 2012, page
  1343. NIH Public Access, 2012.

\bibitem[Zou et~al.(2020)Zou, Wang, Xu, Chen, Zhang, and Gu]{zou2020epidemic}
D.~Zou, L.~Wang, P.~Xu, J.~Chen, W.~Zhang, and Q.~Gu.
\newblock Epidemic model guided machine learning for covid-19 forecasts in the
  united states.
\newblock \emph{medRxiv}, 2020.

\end{thebibliography}
\bibliographystyle{abbrvnat}

\onecolumn
\newpage
\appendix
\section{Graphical Model}
\begin{figure}[ht] 
  \centering
  {\includegraphics[width=100mm]{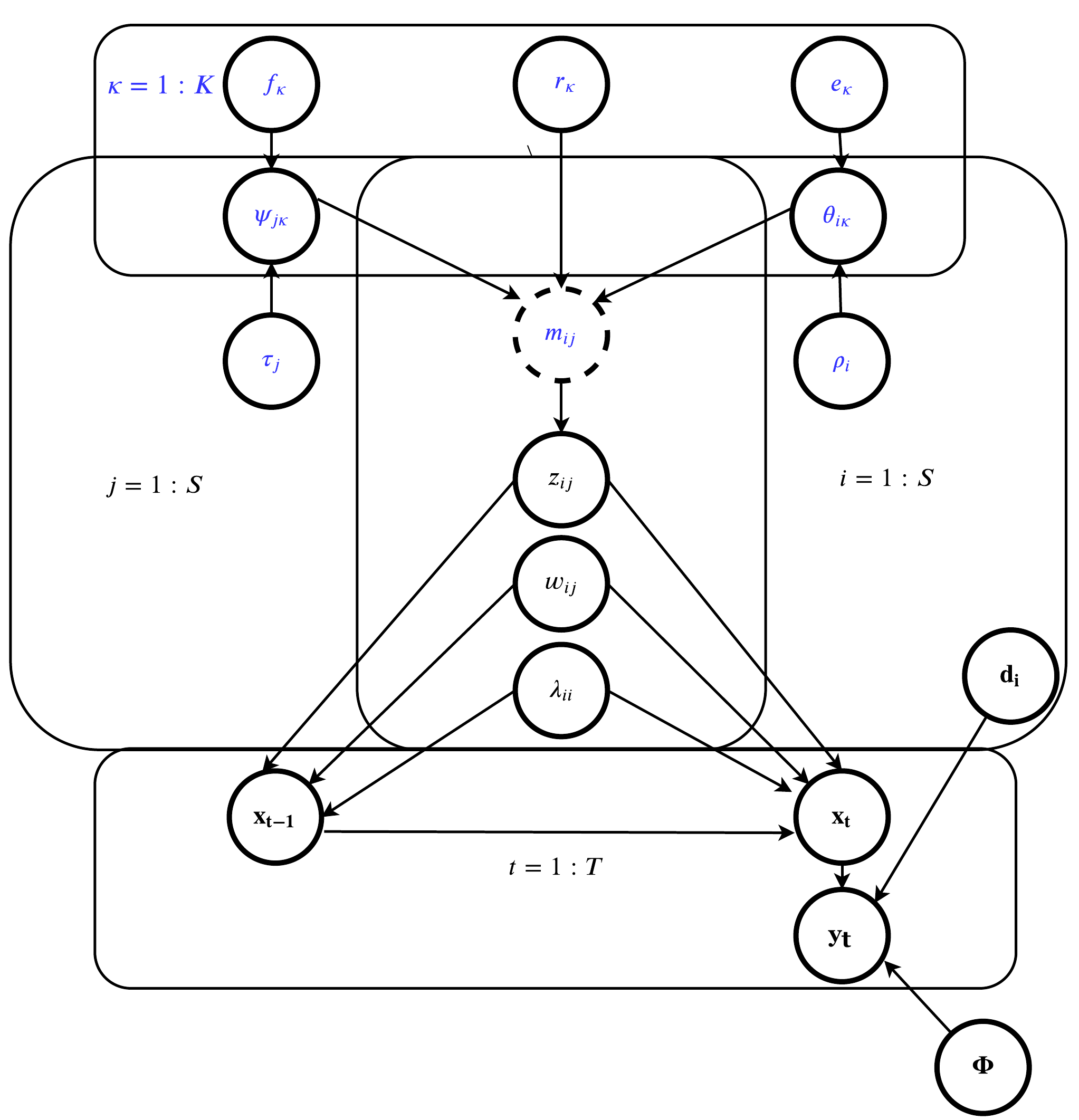}}%
\vspace{0mm}
 
\caption{\small Graphical description of GGP-LDS.}
  \label{fig:graphical_model}
\end{figure}

\section{Proofs}
\label{proof_apx}

\begin{proof}[Proof of Lemma 1]
We first notice $z_{ij\kappa}$ can be equivalently generated under the Bernoulli-Poisson link \citep{epm_aistats2015} as
$$
z_{ij\kappa}=\delta(m_{ijk}\ge 1),~m_{ij\kappa}\sim\mbox{Pois}(r_\kappa\theta_{i\kappa}\psi_{j\kappa}).
$$
As $\sum_{i}\sum_j z_{ij\kappa}\le \sum_i\sum_j m_{ij\kappa}$,  it is sufficient to prove $\E[\sum_i\sum_j m_{ij\kappa}]$ is finite, which is true as $\E[\sum_i\sum_j m_{ij\kappa}] =r_{\kappa}\E[\sum_i\theta_{i\kappa}]\E[\sum_j\psi_{j\kappa}]= r_{\kappa} \gamma_\rho\gamma_\tau/(e f)$.
\end{proof}

\begin{proof}[Proof of Lemma 2]
Following the proof of Lemma 1, with model properties, $\textstyle\Zmat=\delta(\Mmat\ge 1), ~\Mmat\sim\mbox{Pois}(\sum_{k=1}^\infty r_{\kappa} \thetav_{\kappa}\psiv_\kappa^T)$, we have
$\E[\sum_{i}\sum_j z_{ij}]\le \E[\sum_{i}\sum_j \sum_{\kappa}\gamma_\kappa \theta_{i\kappa}\psi_{j\kappa}]=\E[\sum_{\kappa}r_\kappa\gamma_\rho\gamma_\tau/(e f)] =\gamma_0\gamma_\rho\gamma_\tau/(ce f) $.
\end{proof}

Below we show the total energy captured by all non-dynamic states is small: Due to the normal-gamma construction, for state $i$ we have
\begin{ceqn}
\begin{align*}
P(x_{1i},\ldots,x_{Ti} \given a,b) &=\textstyle \int_{0}^\infty \left[\prod_{t=1}^T\mathcal{N}(x_{ti};0,\lambda_i^{-1})\right]\mbox{Gamma}(\lambda_i;a,1/b) \,d\lambda_i \notag\\
&=\textstyle\frac{b^a\Gamma(a+\frac T 2)}{(2\pi)^{\frac T 2} \Gamma(a)} \left(b+\frac{1}{2} \sum_{t=1}^T x^2_{ti}\right)^{-\left(a+\frac T 2 \right)}.\label{eq:normalgamma}
\end{align*}
\end{ceqn}
 Similar to the analysis in \citet{tipping2001sparse}, if we let $a=b=0$, we obtain an improper prior as
$p(\sum_{t=1}^T x_{ti}^2)\propto (\sum_{t=1}^T x_{ti}^2)^{-T/2}$, which is sharply peaked at $\sum_{t=1}^T x_{ti}^2=0$. Thus choosing small $a$ and $b$, the proposed model will penalize   the total energy captured by node (state) $i$, expressed as $\sum_{t=1}^T x^2_{ti}$, if it is a zero-degree node.

\section{Bayesian Inference via Gibbs Sampling}\label{Gibbssampling}
\begin{enumerate}[label=\arabic*)]
\item 
\textbf{\emph{Sample $\omega_{vt}$  for GGD-GLDS}.}
Using data augmentation for negative binomial regression, as in \citet{zhou2012lognormal} and \citet{polson2013bayesian}, %
we denote $\omega_{vt}$ as a random variable drawn from the Polya-Gamma ($\mbox{PG}$) distribution \citep{polson2011default} as
$
\omega_{vt}\sim\mbox{PG}\left(y_{vt}+\eta,~0 \right), %
\notag
$
under which we have 
$\E_{\omega_{vt}} \left[\exp(-\omega_{vt}(\xi_{vt})^2/2)\right] = {\cosh^{-(y_{vt}+\eta)}(\xi_{vt}/2)}$. 
Since $\yv_{t}\sim{\mbox{NB}}(\eta,\sigma(\Dmat \xv_{t}))$, the likelihood of $\xiv_{t}:= \Dmat \xv_{t} $ can be expressed as
\begin{align}
\mathcal{L}(\xiv_{t})&\propto \prod_{v=1}^V\frac{{(e^{\xi_{vt}})}^{y_{vt}}}{{(1+e^{\xi_{vt}})}^{y_{vt}+\eta}} \notag\\
& = \prod_{v=1}^V\frac{2^{-(y_{vt}+\eta)}\exp({\frac{y_{vt}-\eta}{2}\xi_{vt}})}{\cosh^{y_{vt}+\eta}(\xi_{vt}/2)}\nonumber\\ &\propto \prod_{v=1}^V\exp\left({\frac{y_{vt}-\eta}{2}\xi_{vt}}\right)\E_{\omega_{vt}}\left[\exp[-\omega_{vt}(\xi_{vt})^2/2]\right]. \notag
\end{align} 
Combining the likelihood
$
\mathcal{L}(\xi_{vt}, \omega_{vt})\propto \exp\left({\frac{y_{vt}-\eta}{2}\xi_{vt}}\right)\exp[-\omega_{vt}(\xi_{vt})^2/2]
$ and the prior, 
we sample auxiliary PG random variables $\omega_{vt}$ as
\beq
(\omega_{vt}\,|\,-)\sim\mbox{PG}\left(y_{vt}+\eta,~\Dmat(v,:)\xv_t \right),\label{eq:omega_i}
\eeq
where $\Dmat(v,:)$ denotes the $v$th row of $\Dmat$. 
Note to sample from the PG distribution, a fast and accurate approximate sampler of \citet{zhou2016softplus} that  matches the first two moments of the true distribution, with the truncation level set as five, is used in this paper.

\item \textbf{\emph{Sample $\xv_{t}$ for GGP-GLDS.}} Denote $\Omegamat_t = \operatorname{diag}\left(\omega_{1t},\ldots,\omega_{Vt}\right)$. 
Given $\omega_{vt}$, we sample $\xv_t$ as
\ba{({\xv_t}\given-)\sim \mathcal{N}({\muv_t},\Sigmamat_t),}
where, for  if $1\leq t \leq T-1$, we have
 \bas{\bold{\Sigma}_t &=(\bold{D}^T\bold{\Omega}_t\bold{D}+\bold{\Lambda}+(\bold{W} \odot \bold{Z})^{T}\bold{\Lambda}(\bold{W} \odot \bold{Z}))^{-1}, \\
{\muv_t}& = \bold{\Sigma}_t (\Dmat^{T}\frac{\yv_t-\eta}{2}+\bold{\Lambda}(\bold{W} \odot \bold{Z}) \xv_{t-1}+(\bold{W} \odot \bold{Z})^{T}\bold{\Lambda} {\xv_{t+1}}),
}
and 
\bas{
&\bold{\Sigma}_T = (\bold{D}^T\bold{\Omega}_T\bold{D}+\bold{\Lambda})^{-1},\\
&{\muv_T} = \bold{\Sigma}_T \left(\Dmat^{T}\frac{\yv_T-\eta}{2}+\bold{\Lambda}(\bold{W} \odot \bold{Z}){\xv_{T-1}}\right).
}

\item \textbf{\emph{Sample $\xv_{t}$ for GGP-LDS.}} We sample $\xv_t$ as
\ba{({\xv_t}\given-)\sim \mathcal{N}({\muv_t},\Sigmamat_t),}
where, for  if $1\leq t \leq T-1$, we have
 \bas{\bold{\Sigma}_t &=(\bold{D}^T\bold{\Phi}\bold{D}+\bold{\Lambda}+(\bold{W} \odot \bold{Z})^{T}\bold{\Lambda}(\bold{W} \odot \bold{Z}))^{-1}, \\
{\muv_t}& = \bold{\Sigma}_t (\Dmat^{T}\bold{\Phi}\bold{y}_{t}+\bold{\Lambda}(\bold{W} \odot \bold{Z}) \xv_{t-1}+(\bold{W} \odot \bold{Z})^{T}\bold{\Lambda} {\xv_{t+1}}),
}
and 
\bas{
&\bold{\Sigma}_T = (\bold{D}^T\bold{\Phi}\bold{D}+\bold{\Lambda})^{-1},\\
&{\muv_T} = \bold{\Sigma}_T \left(\Dmat^{T}\bold{\Phi}\bold{y}_{T}+\bold{\Lambda}(\bold{W} \odot \bold{Z}){\xv_{T-1}}\right),\\
&\bold{\Sigma_0} = (\bold{H_0}+(\bold{W} \odot \bold{Z})^{T}\bold{\Lambda}(\bold{W} \odot \bold{Z}))^{-1},\\
&{\muv_0} = \bold{\Sigma}_0 (\bold{H_0}{\mv_0}+(\bold{W} \odot \bold{Z})^{T}\bold{\Lambda} {\xv_{1}}).
}

\item \textbf{\emph{Sample $\dv_{s}$.}} We sample $\dv_s$, the $s$th column of $\Dmat$, as
 \ba{ ({\dv_s}\given-)\sim \mathcal{N}(\mv_s,\Emat_s),}
 where for GGP-GLDS, we have
 \bas{
 &\Emat_s = \left(\sum_{t=1}^{T}x_{st}^2\Omegamat_t +\sqrt{V}{\Imat}_{V} \right)^{-1},\\ &\mv_s=\Emat_s\left[\sum_{t=1}^{T}x_{st}\left(\frac{\yv_t-\eta}{2} -\Omegamat_t
 \sum_{i\in\{1,\ldots,S\}\backslash s} \dv_i x_{it}
 \right)\right].
 }
 and for GGP-LDS, we have
 \bas{
 &\Emat_s = \left(\sum_{t=1}^{T}x_{st}^2\Phimat +\sqrt{V}{\Imat}_{V} \right)^{-1},\\ &\mv_s=\Emat_s\Phimat\left[\sum_{t=1}^{T}x_{st}\left({\yv_t} -
 \sum_{i\in\{1,\ldots,S\}\backslash s} \dv_i x_{it}
 \right)\right].
 }

\item  \textbf{\emph{Sample $\eta$ for GGP-GLDS.}}  Using the data augmentation techniques for the NB distribution
\citep{zhou2013negative}, we first sample an auxiliary random variable following the Chinese restaurant table (CRT) distribution and then sample $\eta$ as
\ba{
&(l^{(3)}_{vt}\given-) \sim \mbox{CRT}\left(y_{vt},\eta\right),
\\
&(l^{(4)}\given-) \sim \mbox{CRT}\left(\sum_{v=1}^V\sum_{t=1}^Tl^{(3)}_{vt},\alpha_\eta\right)\\
&(\alpha_\eta \given -)\sim \mbox{Gamma}\Big(\alpha_0+  l^{(4)},\frac{1}{\beta_0-\ln(1-\tilde{p})}\Big),
\\
&(\beta_{\eta}|-)\sim \mbox{Gamma}\Big(\alpha_0+\alpha_\eta,\frac{1}{\beta_0+\eta}\Big),\\
&(\eta\given -)\sim \mbox{Gamma}\Big(\alpha_\eta+\sum_{v=1}^V\sum_{t=1}^{T}l^{(3)}_{vt},\frac{1}{\beta_\eta+\sum_{v=1}^{V}\sum_{t=1}^{T}\zeta_{vt}}\Big),}
where $\zeta_{vt}:= \ln\left(1+e^{ \Dmat(v,:) \xv_{t}} \right)$ and  $\tilde{p}:=\frac{\sum_{v=1}^{V}\sum_{t=1}^{T}\zeta_{vt}}{\beta_\eta+\sum_{v=1}^{V}\sum_{t=1}^{T}\zeta_{vt}}.$

\item \textbf{\emph{Sample $\bold{\Phi}^{-1}$  for GGP-LDS}}. We sample    $\bold{\Phi}^{-1}$ as
\ba{
(\bold{\Phi}^{-1}\given-)\sim \mbox{InverseWishart}(G+\bold {V},V+2+{T}),}
where
$G = G_1-G_2+G_3-G_4$, 
$G_{1}=\sum_{t=1}^{T}\yv_{t}\yv_{t}^{T}$, $G_{2}=\sum_{t=1}^{T}\yv_{t}\xv_{t}^{T}\bold{D}^{T}$,
$ G_{3}=\sum_{t=1}^{T}\bold{Dx_{t}}\xv_{t}^{T}\bold{D}^{T}$, and $G_{4}=\sum_{t=1}^{T}\bold{D}\xv_{t}\bold{y_{t}^{T}}.
$

\item \textbf{\emph{Sample $\Wmat$.}}  We sample $w_{ij}$ as
\ba{
(w_{ij}\given-) \sim \mathcal{N}(\mu_{ij},\tau_{ij}),
}
where
\bas{
 &\tau_{ij}=(z_{ij}\lambda_{i}T_{j}+\varphi_{ij})^{-1}, ~~\mu_{ij}= \tau_{ij}z_{ij}\lambda_{i}Q_{ij},\\
 &T_{j}:=\sum_{t=1}^{T} x_{j(t-1)}^{2},~~ Q_{ij}:=\sum_{t=1}^{T}x_{it}^{-j}x_{j(t-1)},\\ 
 &x_{it}^{-j}:=x_{it}-\sum_{s\in\{1,\ldots,S\}\backslash j} 
 w_{is}z_{is}x_{s(t-1)}.
 } 
 It is noteworthy to mention when $z_{ij} = 0$, it is equivalent to sample $w_{ij}$ from the prior $\mathcal{N}\left(0,\varphi_{ij}^{-1}\right)$. %

\item \textbf{\emph{Sample $\Zmat$.}} We sample $z_{ij}$, the $(i,j)$th element of $\bold Z$, as
\ba{ 
(z_{ij}\given-)\sim \mbox{Bernoulli}[{p_{ij1}}/{(p_{ij1}+p_{ij0}}) ],
}
where 
\bas{
&p_{ij1}:=e^{-\frac{1}{2}(w_{ij}^2T_j\lambda_{i}-2w_{ij}\lambda_{i}Q_{ij})}(1-e^{-\sum_{\kappa}r_{\kappa}\theta_{i\kappa}\psi_{j\kappa}}),\\
&p_{ij0}: = e^{-\sum_{\kappa}r_{\kappa}\theta_{i\kappa}\psi_{j\kappa}}.
}

\item \textbf{\emph{Sample $m_{ij}$.}} We sample augmented count $m_{ij}$ %
as
\ba{
(m_{ij}\given-)\sim z_{ij}\mbox{Pois}_{+}\left(\sum_{\kappa}r_{\kappa}\theta_{i\kappa}\psi_{j\kappa}\right),
}
where $x\sim\mbox{Pois}_{+}(\lambda)$ is a truncated Poisson distribution with $P(x=k)=(1-e^{-\lambda})^{-1}\lambda^ke^{-\lambda}/k!$ for $k\in\{1,2,3,...\}$, which can be efficiently sampled from using a rejection sampler \citep{epm_aistats2015}.

\item \textbf{\emph{Sample $m_{ij\kappa}$.}}
We sample the latent counts, representing how strongly 
a potential transition from state $j$ to state $i$ 
would be associated with the $\kappa^{th}$ community, as
\ba{
(m_{ij1}\hdots m_{ijK}\given-)\sim \mbox{Multinomial}\left(m_{ij},\left[\frac{r_{1}\theta_{i1}\psi_{j1}}{\sum_{\kappa}r_{\kappa}\theta_{i\kappa}\psi_{j\kappa}},\hdots, \frac{r_{K}\theta_{iK}\psi_{jK}}{\sum_{\kappa}r_{\kappa}\theta_{i\kappa}\psi_{j\kappa}} \right]\right).
}

\label{m_ikj}

\item \textbf{\emph{Sample $\theta_{i\kappa}$.}} We sample $\theta_{i\kappa}$ as
\ba{
(\theta_{i\kappa}\given-) \sim \mbox{Gamma}\left(\rho_{i}+\sum_{j=1}^{S}m_{ij\kappa},\frac{1}{e_{\kappa}+r_{\kappa}\sum_{j=1}^{S}\psi_{j\kappa}}\right).
}

\item \textbf{\emph{Sample $\psi_{j\kappa}$.}} We sample $\psi_{j\kappa}$ as
\ba{
(\psi_{j\kappa}\given-) \sim \mbox{Gamma}\left(\tau_{j}+\sum_{i=1}^{S}m_{ij\kappa},\frac{1}{f_{\kappa}+r_{\kappa}\sum_{i=1}^{S}\theta_{i\kappa}}\right).
}

\item \textbf{\emph{Sample $r_{\kappa}$.}} We sample $r_{\kappa}$ as
\ba{
(r_{\kappa}\given-) \sim \mbox{Gamma}\left(\frac{\gamma_{0}}{K}+\sum_{i=1}^{S}\sum_{j=1}^{S}m_{ij\kappa},\frac{1}{c+\sum_{i=1}^S\sum_{j=1}^S\theta_{i\kappa}\psi_{j\kappa }}\right).
}

\item \textbf{\emph{Sample $\rho_{i}$.}} In order to sample $\rho_{i}$, we use a data augmentation technique introduced by  \citet{zhou2013negative} for the negative binomial distribution. Specifically, we augment the model with CRT counts  and then sample $\rho_{i}$ as
\ba{
&(l^{(1)}_{i\kappa}\given-)\sim \mbox{CRT}\left(\sum_{j=1}^{S}m_{ij\kappa},\rho_{i}\right),\\
&(\rho_i\given -)\sim \mbox{Gamma}\left(\frac{\gamma_0}{S}+\sum_{\kappa=1}^{K}l^{(1)}_{i\kappa},\frac{1}{c_\rho-\sum_{\kappa=1}^{K}
\ln(1-p^{(1)}_{i\kappa})
}\right),\\
&p^{(1)}_{i\kappa}:=\frac{r_{\kappa}\sum_{j=1}^{S}\psi_{j\kappa}}{e_{\kappa}+r_{\kappa}\sum_{j=1}^{S}\psi_{j\kappa}}.\notag
}

\item \textbf{\emph{Sample $\tau_{j}$.}} We  sample $\tau_{j}$ as
\ba{
&(l^{(2)}_{j\kappa}\given-)\sim \mbox{CRT}\left(\sum_{i=1}^{S}m_{ij\kappa},\tau_{j}\right)\\
&(\tau_j\given -)\sim \mbox{Gamma}\left(\frac{\gamma_0}{S}+\sum_{\kappa=1}^{K}l^{(2)}_{j\kappa},\frac{1}{c_\tau-\sum_{\kappa=1}^{K}\ln(1-p^{(2)}_{j\kappa})}\right),\\
&p^{(2)}_{j\kappa}:=\frac{r_{\kappa}\sum_{i=1}^{S}\theta_{i\kappa}}{e_{\kappa}+r_{\kappa}\sum_{i=1}^{S}\theta_{i\kappa}}.\notag
}
\item 
\textbf{\emph{Sample $e_{\kappa},f_{\kappa},\lambda_i,\varphi_{ij}$.}}
We sample these parameters as
\ba{
&(e_{\kappa}|-)\sim \mbox{Gamma}\left(\alpha_0+\sum_{i=1}^{S}\rho_i,\frac{1}{\beta_0+\sum_{i=1}^S\theta_{i\kappa}}\right),\\
&(f_{\kappa}|-)\sim \mbox{Gamma}\left(\alpha_0+\sum_{j=1}^{S}\tau_j,\frac{1}{\beta_0+\sum_{j=1}^S\psi_{j\kappa }}\right),\\
&(\lambda_{i}\given-)\sim  \mbox{Gamma}\Big(a+T/2,
\frac{1}{b+\sum_{t=1}^{T}(x_{it}-(W\odot Z)_{i}{\xv_{t-1}})^2/2)}\Big),\\
&(\varphi_{ij}\given-)\sim \mbox{Gamma}\big(\alpha_0+1/2,{1}/{(w_{ij}^2/2+\beta_0)}\big),
}
where $(W\odot Z)_{i}$ is the $i${th} row of $\bold{W}\odot \bold{Z}$.

\end{enumerate}

\newpage

\end{document}